\documentclass[sigplan,nonacm]{acmart}

\AtBeginDocument{%
  }

\usepackage[]{hyperref}
\usepackage{mathtools}
\usepackage{braket}
\usepackage{physics}
\usepackage[normalem]{ulem}
\usepackage{bbm}
\usepackage{graphicx}
\usepackage{tikz}
\usepackage{subcaption}
\usepackage{siunitx}
\usepackage{enumitem}
\usepackage{float}
\usepackage{amsmath}
\usepackage{amssymb, algorithm, algpseudocode}
\newcommand*\circled[1]{\tikz[baseline=(char.base)]{
            \node[shape=circle,draw,inner sep=1pt] (char) {#1};}}

\newenvironment{redtext}{}{} 
\begin{document}
\title{ TreeVQA: A Tree-Structured Execution Framework for Shot Reduction in Variational Quantum Algorithms}

\author{Yuewen Hou}
\email{isaachyw@umich.edu}
\orcid{0009-0004-3035-3197}
\affiliation{%
  \institution{University of Michigan}
  \city{Ann Arbor}
  \state{MI}
  \country{USA}
}

\author{Dhanvi Bharadwaj}
\email{dhanvib@umich.edu}
\orcid{0009-0005-6140-2366}
\affiliation{%
  \institution{University of Michigan}
  \city{Ann Arbor}
  \state{MI}
  \country{USA}
}

\author{Gokul Subramanian Ravi}
\email{gsravi@umich.edu}
\orcid{0000-0002-2334-2682}
\affiliation{%
  \institution{University of Michigan}
  \city{Ann Arbor}
  \state{MI}
  \country{USA}
}

%%%%%%%%%%%%%%% ABSTRACT
\begin{abstract}
Variational Quantum Algorithms (VQAs) are promising for near- and intermediate-term quantum computing, but their execution cost is substantial. Each task requires many iterations and numerous circuits per iteration, and real-world applications often involve multiple tasks, scaling with the precision needed to explore the application’s energy landscape. This demands an enormous number of execution shots, making practical use prohibitively expensive.

We observe that VQA costs can be significantly reduced by exploiting execution similarities across an application’s tasks. Based on this insight, we propose TreeVQA\footnote{TreeVQA is open source at \href{https://github.com/isaachyw/TreeVQA}{https://github.com/isaachyw/TreeVQA}.}, a tree-based execution framework that begins by executing tasks jointly and progressively branches only as their quantum executions diverge.

Implemented as a VQA wrapper, TreeVQA integrates with typical VQA applications. Evaluations on scientific and combinatorial benchmarks show shot count reductions of $25.9\times$ on average and over $100\times$ for large-scale problems at the same target accuracy. The benefits grow further with increasing problem size and precision requirements.

\end{abstract}
%%%%%%%%%%%%%%%

\maketitle % should come after the abstract
%\pagestyle{plain} % should come right after \maketitle - this is not there in the new template

% add the paper content here
\section{Introduction}\label{sec:introduction}

Quantum computers are gradually transitioning from near-term noisy intermediate-scale quantum (NISQ) systems~\cite{preskill2018quantum}, which feature at most a few hundred qubits with limited coherence times and high gate error rates, to intermediate-term Early Fault- Tolerant (EFT) systems~\cite{google_roadmap, iroadmap_2}, expected to host thousands of qubits and employ limited forms of quantum error correction to reduce error rates. These systems will be unable to run long-term quantum applications~\cite{Shor_1997, grover1996fast}, which demand millions of qubits, full fault tolerance, and the ability to execute billions of quantum operations~\cite{O_Gorman_2017}. Despite these limitations, the community remains optimistic that near- and intermediate-term devices can deliver practically useful quantum advantage in key domains such as optimization~\cite{farhi2014quantum}, physics~\cite{kim_evidence_2023a}, and chemistry~\cite{peruzzo2014variational}.

One promising class of algorithms with potential feasibility and utility before full fault tolerance is Variational Quantum Algorithms (VQAs). Their inherent robustness~\cite{VQAsurvey} to noise allows them to deliver useful results even without full-fledged quantum error correction. VQAs have broad applicability, including estimating the energies of molecules~\cite{peruzzo2014variational} and approximating solutions to optimization problems such as MaxCut~\cite{farhi2014quantum}. These hybrid algorithms combine a quantum circuit with parameterized angles (\textit{ansatz}) with a classical optimizer. Through an iterative feedback loop, the VQA explores the solution space and converges to the problem’s ``ground state'', representing the optimal solution.

\begin{figure}[t]
\centering
\includegraphics[width=\columnwidth,trim={0cm 0cm 0cm 0cm},clip]{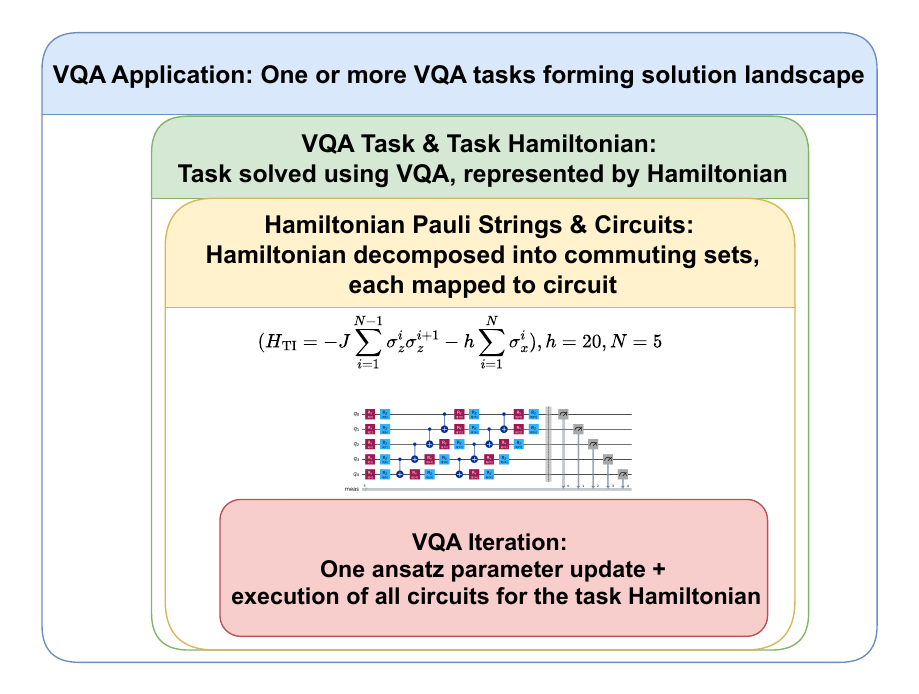}
\caption{Key terminologies and their relations.}
\label{fig:introfig}
\end{figure}

Before delving into VQA details and our proposed work, we introduce key terminology for context: \emph{(i) VQA Application:} An application typically composed of one or more VQA tasks. When multiple tasks are involved, their solutions are combined to construct a solution landscape relevant to the application; \emph{(ii) VQA Task and Task Hamiltonian:} A VQA task is solved using a VQA to find its ground-state solution and is mathematically represented by a Hamiltonian; \emph{(iii) Hamiltonian Pauli String and Circuit:} For a specific VQA task, its Hamiltonian consists of multiple Pauli strings, which can be grouped into sets that fully commute within a set but not across sets. Each set corresponds to a quantum circuit (using the same ansatz but different measurement bases); \emph{(iv) VQA Iteration:} For a specific VQA task, this is a single instance of ansatz parameter updates, as determined by the classical optimizer, followed by the execution of circuits associated with the task Hamiltonian. Fig.\ref{fig:introfig} illustrates this terminology.

\begin{figure*}[t]
\centering
\begin{subfigure}{0.48\textwidth}
    \centering
    \includegraphics[width=\linewidth,trim={0.25cm 0cm 0cm 0cm},clip]{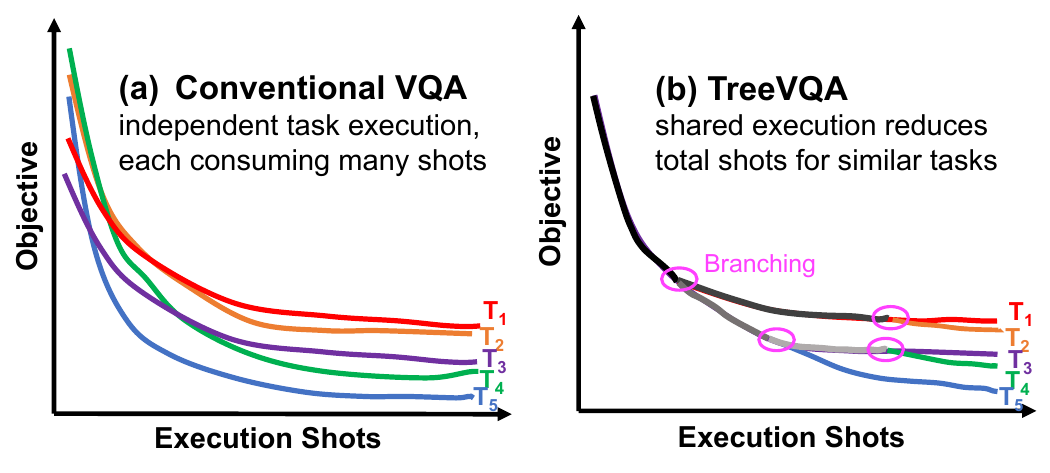}
    \caption{}
    \label{fig:treevqa_intro}
\end{subfigure}
\hfill
\begin{subfigure}{0.48\textwidth}
    \centering
    \includegraphics[width=\linewidth,trim={0cm 0cm 0cm 0cm},clip]{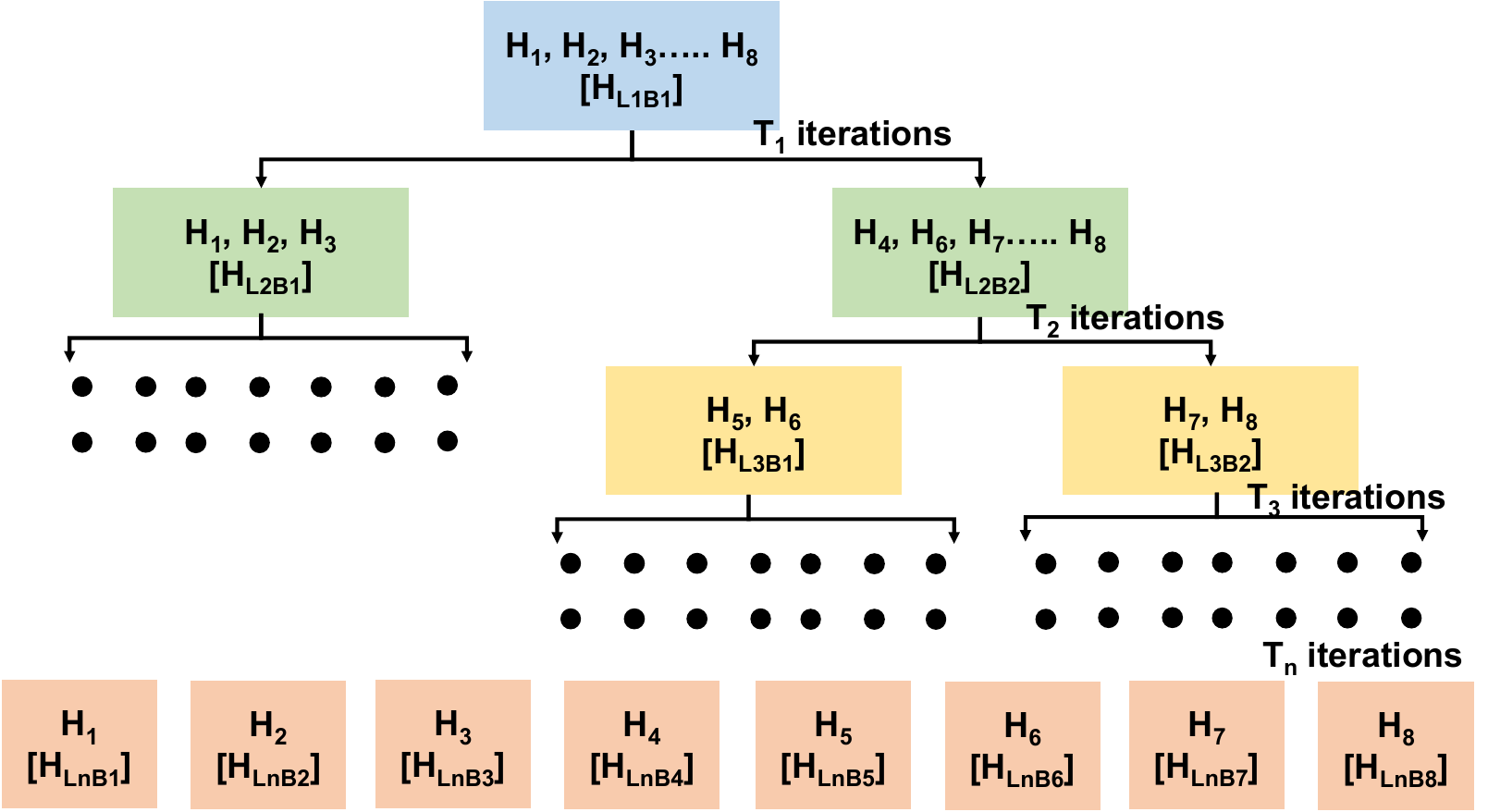}
    \caption{}
    \label{fig:treevqa_overview}
\end{subfigure}
\caption{(a) Comparing the iterative execution of conventional VQA and TreeVQA. (b) TreeVQA strategy overview, branching from the root node to multiple leaf nodes for an eight-VQA task application.}
\label{fig:intro}
\end{figure*}

While VQAs hold promise, their execution faces significant hurdles, the most critical being the enormous number of `shots' (executions) required for any real-world application. This challenge arises from three factors: \circled{a}\ VQA tasks are inherently iterative, often requiring \emph{thousands} of iterations to explore the solution landscape of complex quantum problems, a process further complicated by device noise; \circled{b}\ For practical applications such as chemistry, task Hamiltonians contain \emph{thousands} of Pauli terms, resulting in a large number of circuits per iteration; and \circled{c}\ While the above challenges were associated with individual tasks, real-world applications typically comprise \emph{thousands} of VQA tasks, whose collective ground states define the application’s solution landscape. Together, these factors drive the total number of circuit executions into the \emph{billions} for practical VQA applications, making time and resource costs a show-stopper.

Prior research, both specific to VQA and otherwise, has contributed to both reducing and exacerbating total shot costs, depending on their primary goal. For instance, on the one hand, classical initialization methods~\cite{shaydulin23,shaydulin21, CAFQA_Ravi2022} can help reduce the number of VQA iterations, although their success is often limited and problem dependent. On the other hand, error mitigation strategies such as ZNE~\cite{zne3,giurgica2020digital}, PEC~\cite{Berg2022}, measurement error mitigation~\cite{jigsaw}, and more, all enormously increase the shot counts. The key takeaway is that, despite prior strategies or because of them, the cost of execution shots is still extremely high and deserves attention.

In this work, we make the novel observation that many VQA tasks within an application are highly similar. In particular, the solutions to these tasks (i.e., the quantum states corresponding to their ground states) often exhibit significant overlap. The key idea is that, while these solutions are distinct enough to require individual evaluation for constructing the high-precision solution landscape demanded by the application, their substantial overlaps create an opportunity to share quantum executions across tasks as they iterate toward their respective solutions. By reusing large portions of the VQA process, the total cost of quantum execution can be reduced dramatically.

Based on this insight, we propose \emph{TreeVQA}, a tree-based execution framework that begins by executing all application VQA tasks as a single (or minimal set of) quantum executions. These executions then gradually branch into multiple distinct paths as the quantum states produced by the VQA process converge toward each task’s solution. A comparison of the iterative execution in TreeVQA and conventional VQA is illustrated in Fig.\ref{fig:treevqa_intro}. In TreeVQA, multiple VQA tasks from an application (five in the figure: $T_1$ - $T_5$) begin by executing as a single entity and then eventually diverge. Dissimilar tasks branch out earlier, while similar tasks stick together until closer to convergence. In comparison, the execution cost is much greater for conventional VQA, which executes all of these tasks entirely individually.

The tree execution strategy employed by TreeVQA is more specifically illustrated in Fig.\ref{fig:treevqa_overview}, which shows an example application with eight VQA tasks (with Hamiltonians $H_1$ to $H_8$). Their quantum state evolutions are initially captured by a single representative Hamiltonian, $H_{L1B1}$, which forms the tree root. After $T_1$ VQA iterations, TreeVQA decides to split into two branches based on execution similarity. After the split, $H_{L2B1}$ is the representative Hamiltonian for the first three tasks, and $H_{L2B2}$ is for the remaining five. This branching continues until all tasks eventually execute independently and converge to their respective solutions. The benefit of TreeVQA lies in the shared executions across all internal tree levels prior to the leaf nodes, with the greatest savings occurring near the root. Further details that justify this approach are presented in Section ~\ref{sec:proposal} and beyond. 

\textit{\textbf{The primary contributions and insights are:}}
\begin{enumerate}[leftmargin=*]
    \item We make the novel observation that \emph{substantial reductions in quantum resource costs} are possible by exploiting execution commonalities across multiple related VQA tasks within a single quantum application, an opportunity overlooked in prior work. 
    \item We introduce \emph{TreeVQA}, a hierarchical execution framework that clusters VQA tasks for joint execution and adaptively branches into independent instances only as tasks diverge. This design delivers the best of both worlds: \emph{dramatic quantum resource savings} through shared execution, while \emph{preserving accuracy} through adaptive branching. 
    \item TreeVQA is designed as a \emph{plug-and-play wrapper} that seamlessly integrates with existing VQA algorithms and optimizers, requiring minimal tuning or structural changes. We demonstrate broad applicability through results on VQE and QAOA, using SPSA and COBYLA optimizers. 
    \item Our evaluations show that TreeVQA reduces execution shot counts by \emph{25.9$\times$ on average}, and by \emph{over 100$\times$ in large-scale applications and diverse settings}, all while meeting the same fidelity targets as baseline VQA. 
    \item Crucially, we show that TreeVQA’s benefits \emph{amplify with problem size and precision demands}, making TreeVQA increasingly impactful as quantum applications and hardware scale toward practical use. 
\end{enumerate}

\section{Background and Motivation}\label{section:background}
\subsection{Variational Quantum Algorithms}

Variational quantum algorithms (VQAs) exhibit inherent error resilience due to their hybrid structure, alternating quantum circuits with noise-aware classical optimizers~\cite{peruzzo2014variational, mcclean2016theory}. Prominent VQA applications include the Variational Quantum Eigensolver (VQE)~\cite{peruzzo2014variational}, the Quantum Approximate Optimization Algorithm (QAOA)~\cite{farhi2014quantum}, and Quantum Machine Learning (QML)~\cite{biamonte2017quantum}. This work focuses on VQE and QAOA.
VQE is widely applied in molecular chemistry, condensed matter physics, and quantum many-body systems. QAOA, on the other hand, is tailored for approximate solutions to combinatorial optimization problems, particularly Quadratic Unconstrained Binary Optimization (QUBO).

\begin{figure}[t]
    \centering
    \includegraphics[width=\columnwidth,trim={0cm 0cm 0cm 0cm}]{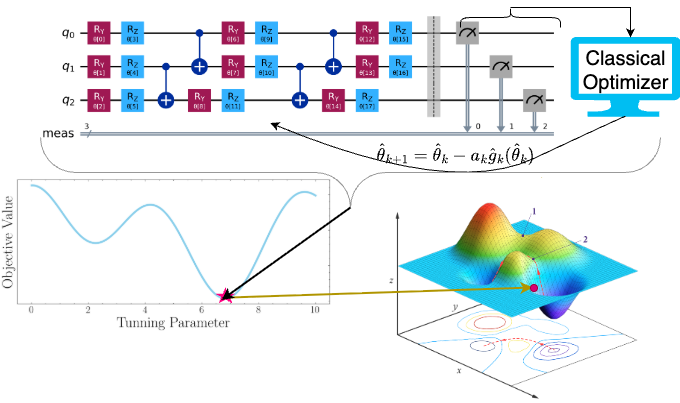}
    \caption{VQA: a hybrid algorithm alternating between classical optimization and quantum execution. Bottom right: Potential Energy Surface for a chemical reaction~\cite{MorschLibreTexts}.}
    \label{fig:vqa}
\end{figure}

Each VQA iteration uses a parameterized quantum circuit, or \textit{ansatz}, which defines the accessible state space and shapes the optimization. The ansatz typically consists of single-qubit rotations and fixed two-qubit gates. A classical optimizer~\cite{9259985, SPSA} iteratively updates the parameters until the objective function converges. This is shown at the top and bottom-left of Fig.\ref{fig:vqa}.
The target problem is encoded as a Hamiltonian, represented as a linear combination of Pauli operators. The goal is to estimate its ground state energy, i.e., the lowest eigenvalue~\cite{mcclean2016theory}. In each iteration, the objective is computed by measuring the Hamiltonian expectation using the current ansatz across appropriate measurement bases.

\subsection{The Cost of Executing a Single VQA Task}
\label{sec:cost_exe_vqa}

VQAs typically demand fewer qubits and shallower circuits compared to many quantum algorithms. However, this comes at the expense of an exceptionally large number of total execution shots to iteratively optimize the ansatz employed to prepare the target quantum state. 

~\cite{gu2021adaptive} highlights that even relatively simple VQE problems in molecular chemistry, such as estimating the ground state energy for $NH_3$ (12 qubits), can require up to $5.5 \times 10^8$ execution shots. This translates to approximately 564 hours of runtime on typical cloud-based quantum devices~\cite{qrio}, incurring monetary expenses on the order of \$5868k, exorbitant costs for fairly modest quantum problems. Scaling to more practically useful applications, ~\cite{VQAsurvey} estimates that a single iteration of 52-qubit VQE on the $Cr_2$ molecule (often considered at the threshold of quantum advantage) could demand 24 years of runtime without parallelization. Alternatively, achieving a runtime of 3ms would require $280 \times 10^9$ qubits under perfect parallelization and ideal overhead assumptions. 

The aforementioned estimates exclude the substantial costs associated with error mitigation techniques such as Probabilistic Error Cancellation (PEC)~\cite{Berg2022} and Zero Noise Extrapolation (ZNE)~\cite{giurgica2020digital}, although these methods are not exclusive to VQAs. 

In a VQA-specific context, ~\cite{dangwal2023varsaw} reports that measurement error mitigation~\cite{jigsaw} can amplify the shot count of a 100-qubit VQA task by approximately 1000x. 
These exorbitant overheads underscore the pressing need for innovative strategies to mitigate resource consumption.

To quantitatively assess VQE cost, we focus on shot count (execution number of quantum circuits). 

The total number of shots depends on the number of Pauli Terms of the decomposed Hamiltonian, the convergence speed of the classical optimizer, and the statistical accuracy desired. 

Given a Hamiltonian expressed as a weighted sum of Pauli terms:
$
H = \sum_{j=1}^{M} c_j P_j, \quad c_j \in \mathbb{R}, \quad P_j \in \{I,X,Y,Z\}^{\otimes n},
$
the expectation value \(\langle H \rangle\) is estimated via measurements. To achieve target accuracy \(\epsilon\), the required number of shots per evaluation is approximately:
$
N_{\text{per-eval}} \approx \frac{\left(\sum_{j=1}^{M} |c_j|\right)^2}{\epsilon^2}.
$
For chemical problems, where the Hamiltonian norm is \(10^2\text{--}10^3\) and energy accuracy target is \(10^{-3}\text{--}10^{-4}\) Hartree, this yields \(10^6\text{--}10^8\) shots per evaluation.

Note that the shot count reflects the cost of a single evaluation. The overall shot cost for the full optimization process, with \(T\) iterations, is given by:
\(
    N_{\text{overall}} = T \times N_{\text{evals-per-iter}} \times N_{\text{per-eval}} 
\).
The iteration number \(T\) is affected by convergence speed, while the number of evaluations per iteration is affected by the classical optimizer.

\begin{figure*}[t]
\centering
\begin{subfigure}{0.34\textwidth}
    \centering
    \includegraphics[width=\linewidth,trim={0cm 0.2cm 2cm 0cm},clip]{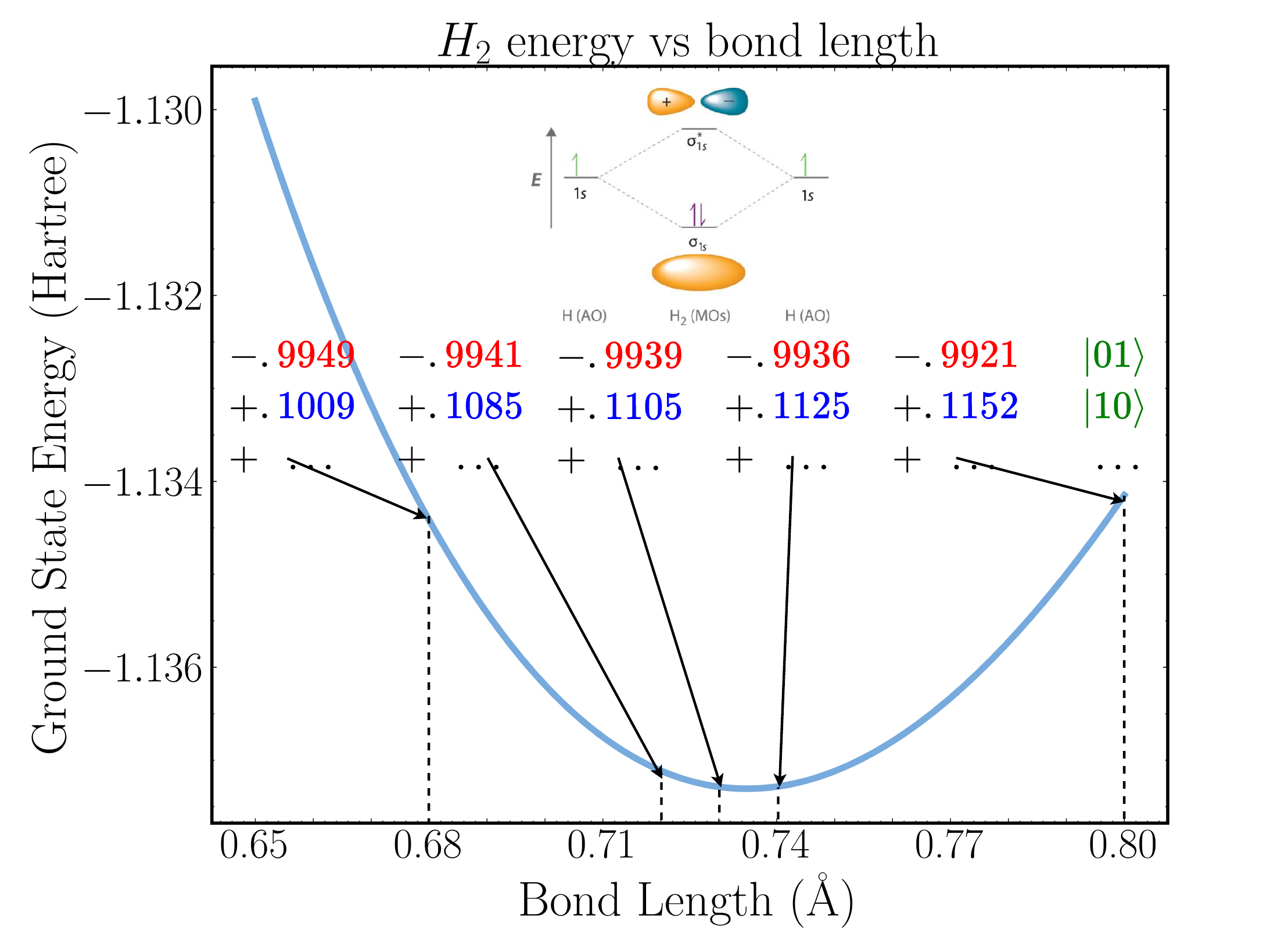}
    \caption{}
    \label{fig:fig1}
\end{subfigure}
\hfill
\begin{subfigure}{0.32\textwidth}
    \centering
    \includegraphics[width=\linewidth,trim={0cm 0cm 0cm 0cm}]{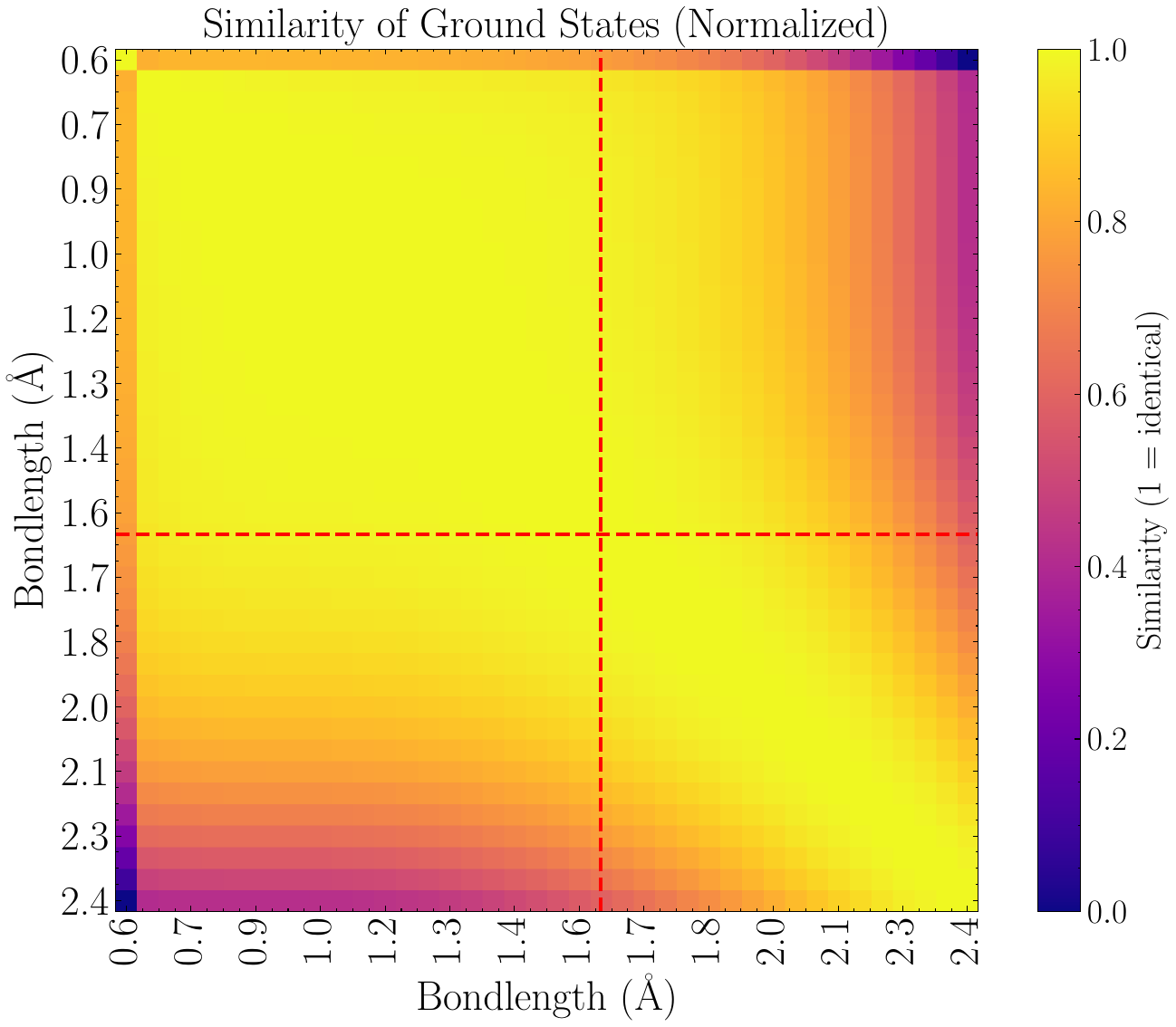}
    \caption{}
    \label{fig:fig2}
\end{subfigure}
\hfill
\begin{subfigure}{0.32\textwidth}
    \centering
    \includegraphics[width=\linewidth,trim={0cm 0cm 0cm 0cm}]{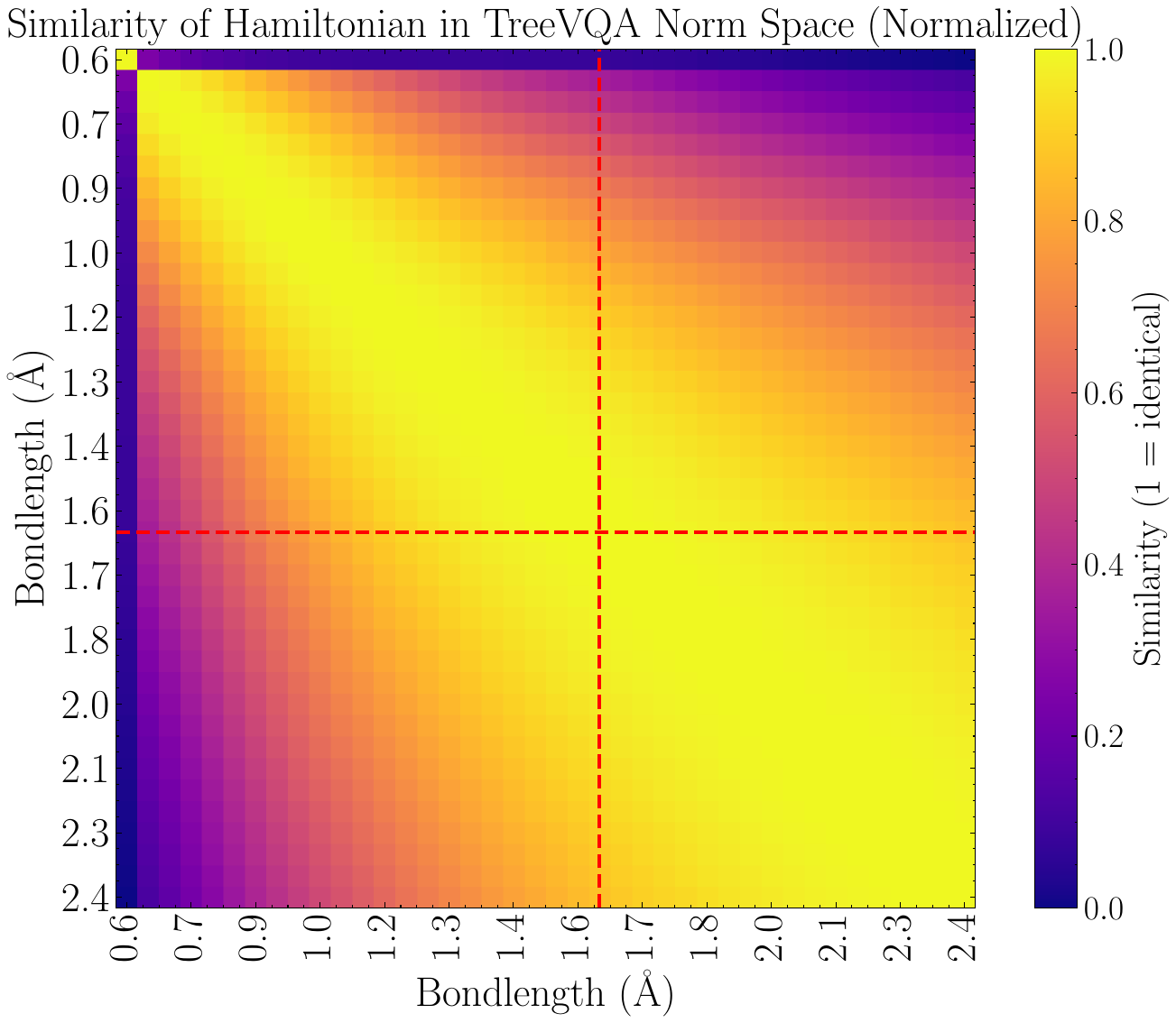}
    \caption{}
    \label{fig:fig3}
\end{subfigure}
\caption{(a) Ground states of $H_2$ and its Molecular Orbital schematic~\cite{LibreTextsPChem} (b) Similarity of ground states of $LiH$ at different bond lengths. (c) Hamiltonian similarity across the VQA tasks, computed through TreeVQA similarity metric (Section \ref{subsec:similarity_matrix}). }
\label{fig:tasksimilar}
\end{figure*}

\subsection{Energy Landscapes from Multiple VQA Tasks}

A primary application of VQAs (specifically VQE) is estimating a molecule’s ground state energy - a task that scales exponentially on classical computers~\cite{Gokhale:2019} and enables predictions of molecular stability, reactivity, and dynamics.
Comparing ground state energies across different molecular configurations is especially critical as it reveals insights about reaction pathways and outcomes. These relationships are captured by the potential energy surface (PES), which maps a molecule’s energy as a function of atomic positions. The PES (shown to the bottom right of Fig.\ref{fig:vqa}) highlights key features such as minima (stable states), maxima (transition states), and saddle points. 
High-accuracy ground state estimates improve the PES resolution, enabling more reliable predictions of molecular stability, vibrational spectra, and dynamic behavior~\cite{adaptvqe}.

Constructing such landscapes with VQAs requires solving multiple VQA tasks, since each molecular configuration corresponds to a separate Hamiltonian with a unique ground state energy. A finer landscape that faithfully captures small variations between nearby points demands both higher precision in each task and a larger number of tasks overall. In the PES shown to the bottom right of Fig.\ref{fig:vqa}, a single point on the energy surface corresponds to a VQA task, and many 1000s of tasks are required to construct the entire landscape.

The concept of constructing detailed energy landscapes via VQA tasks extends to other domains. In condensed matter physics, VQAs can model the electronic ground state to investigate material properties and phase transitions, with energy surfaces shaped by atomic configurations and interaction strengths~\cite{cmp_vqe_PhysRevResearch}.
In protein folding, energy landscapes determine folding pathways through numerous local minima and transition states ~\cite{protein_folding_vqe_robert2021resource}. VQAs can help map these complex landscapes. 

Analogous to VQE in physical systems, constructing `energy' landscapes with QAOA is also highly relevant and involves solving multiple related optimization tasks, each typically corresponding to a graph instance with specific edge weights. These optimization problems can correspond to a wide variety of real-world use cases, such as in weather forecasting, traffic management, supply chain, smart grids (which we will detail later), and more.

\section{Observation: VQA Task Similarity}

The central insight motivating our execution model is the inherent similarity among the many VQA tasks that collectively define an application’s energy landscape.

The theoretical foundation of this approach is rooted in the \textbf{adiabatic theorem}~\cite{Born1928}, which establishes that for a system described by a slowly varying Hamiltonian, the ground state wavefunction evolves continuously. Specifically, if a Hamiltonian $H_1$ is adiabatically transformed into a nearby Hamiltonian $H_2$ in the system's parameter space, the corresponding ground state wavefunction $\psi_1$ will evolve smoothly into $\psi_2$. This adiabatic evolution is contingent upon the persistence of a non-zero energy gap ($\Delta E > 0$) between the ground and first excited states. If this gap were to close, the system would approach a quantum phase transition, permitting non-adiabatic transitions and a potentially discontinuous change in the ground state's character.
This principle is illustrated in \autoref{fig:fig1}, which presents the ground state wavefunction of the H$_2$ molecule at five bond lengths. In this representation, the state is expanded in a basis corresponding to the occupation of molecular orbitals by the two covalent electrons. The figure demonstrates that the coefficients of the primary basis states, such as the covalent configurations $\ket{01}$ and $\ket{10}$, vary very gradually and in close relation to the changes in the internuclear distance.

To further demonstrate that systems with similar configurations (e.g., bond lengths for molecules) exhibit correlated behavior, we construct Fig.~\ref{fig:fig2}, which quantitatively shows the overlap between ground states of VQE tasks at different bond lengths for LiH dissociation. The heatmap places bond length along both axes, with each point $(x, y)$ denoting the overlap between ground states at those lengths. Bright yellow regions correspond to high overlap (evident at small bond length differences), while overlap diminishes at extreme bond lengths.
Complementing this, Fig.\ref{fig:fig3} illustrates that such similarity can be inferred directly from the task Hamiltonians. Specifically, we compare the task Hamiltonians via their non-trivial Pauli coefficients using the $\ell_1$ norm, and compute affinities through a Gaussian kernel (details in Section \ref{subsec:similarity_matrix}). This coefficient-based approach yields consistent insights, confirming that neighboring tasks share structural and spectral similarity.

These observations have direct implications for TreeVQA. Since the VQA's quantum state is prepared by a parameterized unitary transformation, the continuity of the ground state wavefunction implies that the optimal set of variational parameters is also a continuous function of the system's Hamiltonian. The TreeVQA algorithm is designed to capitalize on this insight. It leverages the expected continuity of optimal parameters to guide and accelerate the optimization process for a family of related Hamiltonians. Furthermore, TreeVQA incorporates a dynamic monitoring mechanism capable of detecting signatures of a quantum phase transition, such as abrupt changes in the optimization landscape, allowing it to adapt its strategy and effectively navigate these critical points. 

Finally, although these ideas are motivated here in the context of molecular chemistry, the underlying principle generalizes across domains. 

\section{TreeVQA: Tree-based VQA Execution}
\label{sec:proposal}

TreeVQA is a hierarchical framework designed to efficiently handle applications involving multiple variational quantum algorithm (VQA) tasks with underlying similarity. By clustering related tasks, TreeVQA adaptively optimizes shared circuit parameters across groups, thereby reducing the total number of quantum measurements. As optimization proceeds, clusters are recursively subdivided, enabling increasingly fine-grained adaptation to each Hamiltonian.
Fig.~\ref{fig:treevqa_overview} illustrates this process for an application with eight VQA tasks, each associated with a distinct Hamiltonian ($H_1$ through $H_8$). For example, these Hamiltonians could represent a molecule evaluated at eight different geometries (e.g., bond lengths), where each $H_i$ corresponds to the molecular Hamiltonian at a specific geometry. The objective is then to find the ground state at each geometry.
TreeVQA execution is detailed below:

\begin{enumerate}[leftmargin=*]
    \item To begin, all VQA tasks are initially grouped into a single cluster, as they may share a suitable reference quantum state (note: this is not required in TreeVQA - one could also start with multiple clusters if desired). Each cluster corresponds to a single ``mixed" Hamiltonian that represents all task Hamiltonians within it (see Section~\ref{sec:superset} for details). In~\autoref{fig:treevqa_overview}, this is shown in blue as $H_{L1B1}$ (Level1, Branch1). Variational optimization is then performed on an ansatz, with its parameters updated based on the expectation value of the mixed Hamiltonian. This single-cluster execution continues until branching events occur, as described next.
    
    \item After a warmup phase, the optimization of each VQA cluster is monitored using a sliding window method that evaluates slopes (see Sections~\ref{sec:optimization}). When a split condition is triggered, signaling either stagnation in optimization or divergence among Hamiltonians within the cluster, the cluster is divided. The split is guided by precomputed task similarity information (see Section~\ref{sec:split}). In~\autoref{fig:treevqa_overview}, after $T_1$ iterations of single-cluster execution, the process branches into two clusters, shown in green.
    
    \item The splitting process employs similarity-based clustering (see Section~\ref{sec:clustsplit}) to identify natural groupings of Hamiltonians. In the example, the first three tasks exhibit similar characteristics and are grouped into one cluster with mixed Hamiltonian $H_{L2B1}$ (Level 2, Branch 1), while the remaining five tasks form a second cluster with mixed Hamiltonian $H_{L2B2}$ (Level 2, Branch 2). Each new cluster inherits the parameters of its parent, enabling warm-started optimization.
    
    \item This hierarchical splitting process continues recursively; for instance, the second branch may further divide into two level-three branches (yellow). Each split occurs only when the specified conditions are met, ensuring efficient utilization of quantum resources. The procedure terminates either when the total shot budget is depleted or when convergence is achieved (see Section~\ref{sec:central}). Additionally, post-processing steps are applied to refine the results (not shown in the figure, see Section~\ref{sec:post_processing}).
\end{enumerate}

\begin{figure}[t]
    \centering
    \includegraphics[width=\columnwidth,trim={10cm 0cm 7cm 0cm},clip]{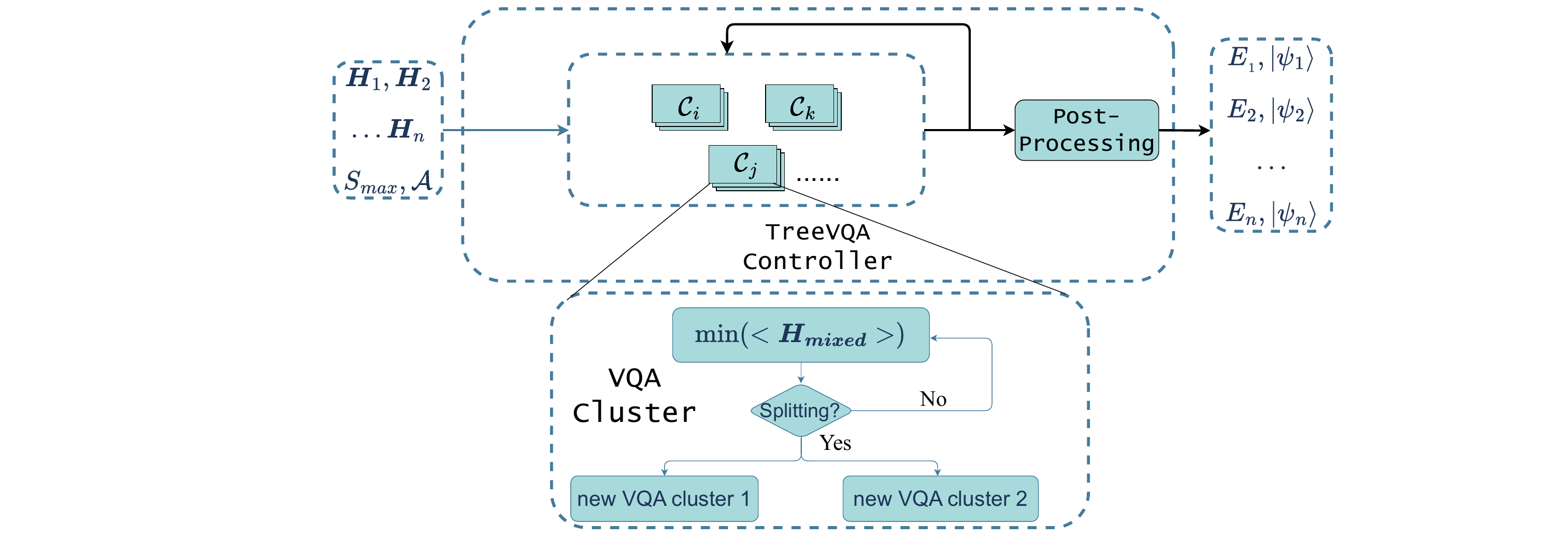}
    \caption{TreeVQA: A central controller and VQA clusters}
    \label{fig:arch}
\end{figure}

\section{TreeVQA Framework Architecture}\label{sec:design}

The TreeVQA framework is designed as a central controller to orchestrate the execution and communication among the VQA clusters. We discuss these next.

\subsection{TreeVQA Central Controller}
\label{sec:central}

The Central Controller orchestrates the overall execution of the algorithm, including cluster creation, splitting logic, resource coordination, and enforcement of the global shot budget, as illustrated in \autoref{fig:arch}. The controller workflow is detailed in Algorithm~\autoref{alg:1}.

For a given application, TreeVQA receives the set of task Hamiltonians and their initial states (e.g., Hartree–Fock, CAFQA, or Red-QAOA~\cite{szabo1996modern}; see \autoref{sec:evaluation} for details). Based on the number of unique initial states, TreeVQA initializes a corresponding number of VQA clusters, each containing Hamiltonians sharing the same initial state. The controller manages the entire process: each VQA cluster optimizes its assigned Hamiltonians and may split into child clusters as necessary (see \autoref{sec:cluster}), with child clusters inheriting parameters from their parent. After all iterations, a final post-processing step evaluates each Hamiltonian across all cluster states to select the optimal solution (see \autoref{sec:post_processing}).

\begin{algorithm}[h]
\caption{TreeVQA Global Controller}
\begin{algorithmic}[1]
\State \textbf{Input:} Hamiltonians $\{H_1,\ldots,H_N\}$, total shot budget $S_{\max}$, ansatz $\mathcal{A}$
\State Initialize root cluster $\mathcal{C}_0 \gets \{H_1, \ldots, H_N\}$
\State $\mathbb{C} \gets \{\mathcal{C}_0\}$, \quad $S_{\text{total}} \gets 0$
\While{$S_{\text{total}} < S_{\max}$}
    \For{\textbf{each} cluster $\mathcal{C} \in \mathbb{C}$}
        \State \textsc{VQA-Cluster-Step}($\mathcal{C}$) \Comment{See Algorithm~\ref{alg:vqa_cluster}}
        \State $S_{\text{total}} \gets S_{\text{total}} + S_{\mathcal{C}}$ \Comment{Cluster shot usage}
    \EndFor
    \State \textbf{Update} $\mathbb{C}$ by replacing split clusters with their children
\EndWhile
\State $\mathbb{C}_{\text{final}} \gets \mathbb{C}$ \Comment{Final cluster set}
\For{\textbf{each} Hamiltonian $H_i$}
    \For{\textbf{each} cluster $\mathcal{C} \in \mathbb{C}_{\text{final}}$}
        \State $E_i^{\mathcal{C}} \gets \langle \psi(\theta_{\mathcal{C}}) | H_i | \psi(\theta_{\mathcal{C}}) \rangle$ \Comment{Post-processing evaluation}
    \EndFor
    \State $\mathcal{C}_{\text{best}}(H_i) \gets \operatorname{argmin}_{\mathcal{C} \in \mathbb{C}_{\text{final}}} E_i^{\mathcal{C}}$ \Comment{Best state for each Hamiltonian}
\EndFor
\State \textbf{Return:} For each $H_i$, the best energy $E_i^{\mathcal{C}_{\text{best}}}$ and corresponding state $\psi(\theta_{\mathcal{C}_{\text{best}}(H_i)})$
\end{algorithmic}
\label{alg:1}
\end{algorithm}

\subsection{VQA Cluster Functionality}
\label{sec:cluster}

A \emph{VQA Cluster} is a fundamental computational unit in the TreeVQA framework (see zoomed region in \autoref{fig:arch}), responsible for jointly optimizing a shared parameterized quantum state \( |\psi(\theta)\rangle \) over a subset of Hamiltonians \( \{H_1, H_2, \dots, H_N\} \). By aggregating the Pauli terms from its assigned Hamiltonians, the cluster constructs a mixed Hamiltonian and performs collective parameter optimization to approximate the ground states of all members, leveraging their similarity. If the optimization trajectories of certain Hamiltonians diverge significantly (quantified by a similarity metric) the cluster adaptively splits into child clusters to maintain optimization efficiency and accuracy. We detail the functionality of the VQA Cluster and the design philosophy behind it in the following subsections.

\begin{algorithm}[h]
\caption{VQA-Cluster-Step}
\label{alg:vqa_cluster}
\begin{algorithmic}[1]
\State Compute the mixed Hamiltonian \(H_{\text{mixed}}\)
\State Compute similarity matrix \(S_\mathcal{C}\)
\State Initialize parameters \(\theta_\mathcal{C}\) (inherited from parent)
\While{true}
\State Optimize \(\theta_\mathcal{C}\) over the mixed Hamiltonian.
\State Track loss values for the mixed and individual Hamiltonians.
\If{warmup period is complete}
    \State Compute \(\text{slope}_{\text{mixed}}\) via linear regression on mixed loss
    \State Compute \(\text{slope}_i\) for each \(H_i \in \mathbb{H}_\mathcal{C}\)
\EndIf
\If{\(\text{slope}_{\text{mixed}} < \epsilon_{\text{split}}\) \textbf{or} any \(\text{slope}_i > 0\)}
    \State Split \(\mathcal{C}\) into \(\mathcal{C}_1\), \(\mathcal{C}_2\) via spectral clustering on \(S_\mathcal{C}\)
    \State Inherit parameters: 
    \(\theta_{\mathcal{C}_1}=\theta_{\mathcal{C}_2}  = \theta_\mathcal{C} \)
    \State Mark \(\mathcal{C}\) as retired
    \State \Return \(\mathcal{C}_1, \mathcal{C}_2\)
\EndIf
\EndWhile
\State \textbf{Record} shot usage \(S_\mathcal{C}\)
\end{algorithmic}
\end{algorithm}

\subsubsection{Cluster Mixed Hamiltonian Construction}
\label{sec:superset}
When the target application's Hamiltonians vary over its tasks, small but nonzero integrals can emerge or vanish \cite{pyscf}, resulting in new or missing Pauli terms in its qubit Hamiltonian. To handle multiple Hamiltonians consistently, the VQA cluster first identifies the superset of all Pauli terms:
$
\mathcal{P} = \bigcup_{i=1}^N \{P_j \mid P_j \in H_i\}, \quad |\mathcal{P}| = M.
$
Then, for each Hamiltonian \( H_i \), it pads any missing terms in \( \mathcal{P} \) with zero. Since many of these extra terms arise from integrals that are either zero or negligibly small, the padding is typically minimal, and any new coefficients are close to zero. Thus, while padding ensures consistency across VQA tasks, it does not significantly increase the Hamiltonian's complexity or alter its underlying characteristics.

The cluster's mixed Hamiltonian (which represents the set of Hamiltonians this cluster will handle) is given by:
\(
H_{\text{mixed}} = \frac{1}{N} \sum_{i=1}^N H_i^{\text{padded}} = \sum_{k=1}^M \left( \frac{1}{N} \sum_{i=1}^N c_{ik} \right) P_k.
\)
This construction yields a Hermitian operator according to the linearity of quantum mechanics, incurs minimal computational cost, and provides a representative optimization landscape for all cluster members.
The corresponding objective function is:
\(
\mathcal{L}(\theta) = \langle \psi(\theta) | H_{\text{mixed}} | \psi(\theta) \rangle = \frac{1}{N} \sum_{i=1}^N \langle \psi(\theta) | H_i^{\text{padded}} | \psi(\theta) \rangle.
\)

\subsubsection{Iterative VQA Optimization and Sliding Window Slope Monitoring}
\label{sec:optimization}

After the preliminary step, VQA optimization is performed on each cluster's mixed Hamiltonian, iteratively updating the cluster's ansatz parameters \( \theta \) using the selected VQA optimizer. In this work, we focus on the Simultaneous Perturbation Stochastic Approximation (SPSA) optimizer \cite{SPSA}, though the framework can be generalized to other optimizers. The parameter update is given by:
$
\theta_{t+1} = \theta_t - \eta_t \, \frac{\mathcal{L}(\theta_t + \Delta_t) - \mathcal{L}(\theta_t - \Delta_t)}{2\,\Delta_t},
$
where \( \Delta_t \) is a random perturbation vector and \( \eta_t \) is the learning rate.

After \( T_{\text{warmup}} \) steps, allowing the optimizer to freely explore the parameter space without splitting constraints, we compute a running slope over a window of \( W \) loss values:
\(
\text{slope}_t = \text{LinearRegression}\left(\{\mathcal{L}(\theta_{t-W+1}), \dots, \mathcal{L}(\theta_t)\}\right).
\)
For each Hamiltonian \( H_i \) in the cluster, we also track $\text{slope}_{i,t}$ after each iteration in the same way, where \( \mathcal{L}_i(\theta) = \langle \psi(\theta) | H_i | \psi(\theta) \rangle \). Notably, this individual tracking incurs only a linear classical computational cost, not a quantum execution cost. This tracking is used in the next subsection.

\subsubsection{Cluster Splitting}
\label{sec:split}

The tracking described above allows for cluster splitting when appropriate, ensuring that each VQA task reaches its own target. We trigger the splitting if either of the following conditions is met:
$
|\text{slope}_t| < \epsilon_{\text{split}} 
$
or
$
\exists\, i \;, \text{such that} \; \text{slope}_{i,t} > 0 
$
These conditions capture both a stalled overall optimization and an unproductive path for any individual Hamiltonian within the cluster.

If splitting is not triggered, the optimization will continue until retirement. However, when splitting occurs, it indicates that the cluster has determined that some Hamiltonians are too distant from others to be optimized together. As a result, the cluster will be divided into several new clusters. The key challenge is to identify which Hamiltonians are "far" from the others and which are "close," i.e., to quantify distances between Hamiltonians. Moreover, the distance should capture both gradient direction and ground-state energy shift while remaining efficient to compute; we discuss this next.

\subsubsection{Hamiltonian Similarity: Hilbert Space Distance and Gradient Alignment}
\label{subsec:similarity_matrix}

To quantify the similarity between Hamiltonians \( H_i \) and \( H_j \), we compute the \(\ell_1\) norm between their padded coefficient vectors:

\(
\mathbf{c}_i = (c_{i1}, \dots, c_{iM}), \quad
d(H_i, H_j) = \|\mathbf{c}_i - \mathbf{c}_j\|_1 = \sum_{k=1}^M |c_{ik} - c_{jk}|.
\)
This metric is computationally efficient and serves as a practical proxy for Hilbert space proximity, as the operator norm difference between two Hamiltonians is bounded by their \(\ell_1\) coefficient distance:

$
\|H_i - H_j\|_{\mathrm{op}} = \left\|\sum_{k=1}^M (c_{ik} - c_{jk}) P_k \right\|_{\mathrm{op}} \leq d(H_i, H_j).
$
According to perturbation theory~\cite{Sakurai_Napolitano_2020}, if \(\|H_i - H_j\|_{\mathrm{op}} \ll \Delta\) (where \(\Delta\) is the spectral gap), the ground-state energies and eigenstates of \(H_i\) and \(H_j\) remain close. Thus, a small \(\ell_1\) distance indicates similar ground-state properties and optimization landscapes.

To facilitate smooth and normalized pairwise similarity, we construct a Gaussian (RBF) kernel:
$
S_{ij} = \exp\left(-\frac{d(H_i, H_j)^2}{2\sigma^2}\right),
$
where \(\sigma\) is set to the median pairwise distance in the dataset. The resulting similarity matrix \(S \in \mathbb{R}^{N \times N}\) guides cluster splitting.
Beyond static similarity, the \(\ell_1\) distance also reflects the alignment of gradient directions during VQA optimization. The gradient of the loss function \(\langle \psi(\theta) | H_i | \psi(\theta) \rangle\) with respect to parameters \(\theta\) is determined by the dominant Pauli coefficients in \(H_i\). Hamiltonians with similar coefficient vectors yield aligned gradients, enabling effective joint optimization within a cluster. Conversely, a Hamiltonian with a distinct coefficient vector will have a divergent gradient direction, impeding convergence if grouped with dissimilar tasks. The similarity matrix \(S\) thus not only encodes Hilbert space proximity but also captures the coherence of optimization trajectories.
When the similarity matrix is available, spectral clustering~\cite{vonluxburg2007tutorialspectralclustering} is employed to partition the cluster when the splitting condition is met, as shown below.

\subsubsection{Cluster Assignment via Spectral Clustering}
\label{sec:clustsplit}

Upon triggering a cluster split, we apply spectral clustering to the similarity matrix \(S\) to partition the Hamiltonians into  homogeneous subgroups. Spectral clustering constructs the normalized Laplacian of \(S\), computes its leading eigenvectors, and performs k-means clustering in this reduced space to assign each Hamiltonian to one of two child clusters~\cite{vonluxburg2007tutorialspectralclustering}. This approach efficiently identifies natural divisions in the similarity structure, ensuring that Hamiltonians with greater similarity are grouped together.
Each child cluster inherits the parent’s parameter values, enabling warm-started optimization and continuity in the variational process.

\subsection{Post-Processing}
\label{sec:post_processing}

After reaching the shot budget limit (\( S_{\text{total}} > S_{\text{max}} \)), TreeVQA performs  post-processing to ensure optimal results for each Hamiltonian. At this stage, we have a set of final clusters \( \mathbb{C}_{\text{final}} \), each with its own optimized quantum state \( |\psi(\theta_{\mathcal{C}})\rangle \) as shown in Algorithm~\autoref{alg:1}.
For each individual Hamiltonian \( H_i \), we evaluate its expectation value on \textit{all} optimized states from the final clusters. We then select the best state for each Hamiltonian based on these expectation values.
The post-processing approach is computationally efficient because each cluster already logs the expectation values of individual Pauli terms during optimization. Evaluating a Hamiltonian on states from different clusters is essentially a classical recombination of stored values with different coefficients.

\section{Tailoring for QAOA}\label{sec:tailor_for_qaoa}

TreeVQA is broadly suited for all VQAs, but its implementation is more directly suited to VQE. In this section, we discuss tailoring it for QAOA. 

The QAOA ansatz consists of a circuit with $p$ alternating parameterized layers acting on an initial state $\ket{\Psi_0}$. Each layer $\ell$ applies the phasing operator $e^{-i \gamma_{\ell}C}$, adding a complex phase to each bitstring $x$ based on its cost $c(x)$ from the QUBO and a real-valued parameter $\gamma_{\ell}$. This phasing step is applicable to any QUBO \cite{glover19}. The mixing layer, $e^{-i \beta_{\ell}B}$, combines quantum state amplitudes, with the coefficient depending on the Hamming distance between states and a real-valued parameter $\beta_{\ell}$. Specifically, $B = \sum_i X_i$, so $e^{-i \beta_{\ell}B}$ can be implemented as single-qubit $R_x$ rotations on each qubit. The goal is to optimize the $2p$ parameters $(\vec{\gamma}, \vec{\beta})$ to maximize quantum amplitudes for optimal or near-optimal bitstrings $x$.
The performance of QAOA depends on the choice of parameters $(\vec{\gamma}, \vec{\beta})$, and optimizing these parameters can be highly non-trivial~\cite{wang18, farhi22, barak15, hastings19, marwaha21, chou22, lin16, farhi15, hadfield19, streif20}. A common approach is variational optimization, where a quantum computer estimates the expectation value of the quantum state induced by QAOA with parameters $(\vec{\gamma}, \vec{\beta})$ with respect to the cost operator $C$. A classical computer then updates $(\vec{\gamma}, \vec{\beta})$ using this estimate, along with previous ones, and the quantum-classical loop continues until a convergence criterion is met.

The standard QAOA ansatz is problem-specific and minimally parameterized, typically using only $2p$ parameters. To implement TreeVQA effectively, we adopt a circuit ansatz that generalizes across problem instances, enabling the representation of problems with shared structure. For finer control in splitting during TreeVQA, we increase the number of classical parameters per QAOA layer. We achieve this by implementing the multi-angle QAOA (ma-QAOA) \cite{herrman2021maqaoa}, which enhances the expressivity of the traditional structure by assigning individual parameters to each term in the cost and mixer operators, instead of using a single angle for each. Specifically:
$
U(C, \vec{\gamma_{\ell}}) = e^{-i \sum_{a=1}^{m} C_{a} \gamma_{\ell,a}} = 
\prod_{a=1}^{m} e^{-i C_{a} \gamma_{\ell,a}}
$, and 
$
U(B, \vec{\beta_{\ell}}) = e^{-i \sum_{b=1}^{n} B_{b} \beta_{\ell,b}} = 
\prod_{b=1}^{n} e^{-i B_{b} \beta_{\ell,b}}.
$
Here, $m$ is the number of clauses and $n$ is the number of qubits. The total number of parameters becomes $(m+n)p$. The standard QAOA is a special case of ma-QAOA, where all parameters within a cost or mixer layer are identical. ma-QAOA expands the parameter space, allowing TreeVQA to explore more complex solution landscapes and improve optimization efficiency.

\section{Methodology}
\label{sec:methodology}

\subsection{Benchmarks}
\label{subsec:benchmarks}

\begin{table}[thb]
\caption{Chemistry Benchmarks}
\label{tab:cb}
\resizebox{\columnwidth}{!}{%
\begin{tabular}{|l|l|l|l|l|l|}
\hline
                                                                       & $H_2$       & $LiH$      & $BeH_2$     & $HF$       & $C_2H_2^{\ast}$    \\ \hline
\# Ham. terms                                                     & 15       & 496      & 810      & 631      & 5945        \\ \hline
Qubit numbers                                                          & 4        & 12       & 14       & 12       & 28        \\ \hline
Bond range ($\si{\angstrom}$)                                                   & .74-.83 & 1.4-1.7 & 1.2-1.47 & .83-1.1 & 1.15-1.25 \\ \hline
\begin{tabular}[c]{@{}l@{}}Eq. bond ($\si{\angstrom}$)\end{tabular} & .741    & 1.595    & 1.333    & .917    & 1.2  \\ \hline
\end{tabular}%
}
\end{table}

\textbf{Chemistry Benchmarks:}
We select \( H_2 \), \( \text{BeH}_2 \), \( \text{LiH} \), \( \text{HF} \), and \( C_2H_2 \) as primary chemical benchmarks. Using Qiskit’s Jordan--Wigner mapper \cite{tranter2015bravyi}, we convert the fermionic Hamiltonian in the STO-3G basis to a qubit Hamiltonian, with the resulting terms and qubit counts shown in Table \ref{tab:cb}. All molecular spin orbitals are left open \cite{the_qiskit_nature_developers_and_contrib_2023_7828768}, enabling TreeVQA to distinguish configurations near equilibrium.
We initialize both TreeVQA and the baseline to start with the Hartree-Fock state unless specified otherwise.
Each benchmark includes 10 bond-length instances spaced by \( 0.03 \, \text{\AA} \). For \( H_2 \), we use the UCCSD ansatz with 5 instances due to its smaller size. TreeVQA is also applicable to a broader range of molecule geometries.

\noindent\textbf{Physics Benchmarks:}
\label{subsubsec:ising_bench}
We evaluate two spin-\(\frac{1}{2}\) models: (1) The Heisenberg XXZ chain with Hamiltonian \(H_{\text{XXZ}} = J \sum_{i=1}^{N-1} ( \sigma_x^i \sigma_x^{i+1} + \sigma_y^i \sigma_y^{i+1} + \Delta \sigma_z^i \sigma_z^{i+1} )\), where \(\Delta\) is the anisotropy parameter that drives transitions between gapless (\(|\Delta| < 1\)) and gapped phases, with a BKT transition\cite{BKT_transition} at \(\Delta = 1\); and (2) The transverse Ising model with \(H_{\text{TI}} = -J \sum_{i=1}^{N-1} \sigma_z^i \sigma_z^{i+1} - h \sum_{i=1}^{N} \sigma_x^i\), which exhibits a quantum phase transition at \(h = J\) between ordered and disordered phases\cite{Fradkin_2013}. For both models, we set \(J = 1\) and use Qiskit's LinearMapper\cite{qiskitnature2023linearmapper} for spin-to-qubit mapping.

\noindent\textbf{QAOA Benchmark:}
\label{subsubsec:qaoa_benchmark}
The Maximum Cut (MaxCut) is a well-studied NP-hard problem in graph theory, widely used as a benchmark for QAOA. It has applications in statistical physics and circuit design, and involves partitioning the vertices of a graph into two complementary subsets to maximize the sum of edge weights crossing the cut. The cost Hamiltonian for a graph \( G \) with edge set \( E(G) \) is
\(
H_{\text{C}} = \sum_{(i,j) \in E(G)} \tfrac{1}{2} w_{ij} \left( I - Z_i Z_j \right),
\)
where \( w_{ij} \) are edge weights.  
We benchmark TreeVQA on MaxCut as it enables exploration of diverse graph structures under uniform optimization settings. In particular, we evaluate graph instances derived from the IEEE 14-bus test system~\cite{ieee14bus}, a canonical dataset in power systems.  Solving MaxCut on these graphs identifies optimal partitions that maximize power transfer while minimizing inter-partition dependencies; this is key for system reliability and distributed energy coordination in smart grids~\cite{data_driven_qaoa}. In these graphs, buses correspond to vertices and transmission lines/transformers to edges, yielding a weighted graph well-suited for QAOA. We construct isomorphic graph representations where nodes denote buses and weighted edges capture line capacities influenced by load conditions.  By scaling load values, we vary edge weights to simulate operating regimes from light to heavy demand.

\subsection{Evaluation Metrics}
\label{subsec:metric}
We define each benchmark instance, such as molecular configurations at different bond lengths or physical models at varying parameters, as a \emph{task}. For each task $i$, with $E_{gs_i}$ denoting the ground-state energy, the error is defined as $\epsilon_{i} = \frac{E_{gs_i} - E_{i}}{E_{gs_i}}$, with fidelity $F_{i} = 1 - \epsilon_{i}$. Since TreeVQA collectively solves multiple tasks for an application, we aggregate error and fidelity across the $N$ subproblems. We define the fidelity threshold $T$:
$
\forall F_{i}, i \in \{1,2,\dots,N\}, \quad F_{i} \geq T.
$

\begin{figure*}[t]
    \centering
    \includegraphics[width=\textwidth]{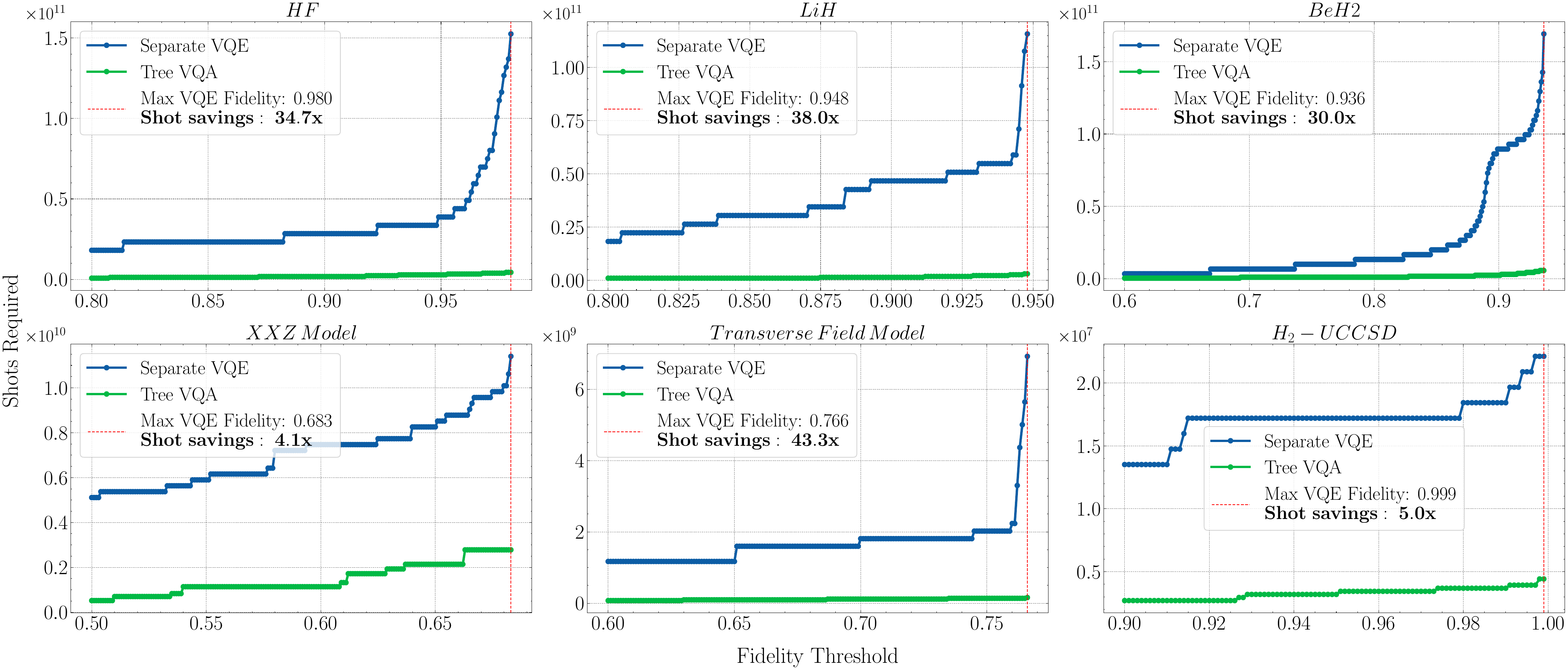}
    \caption{Shot reduction for TreeVQA compared to the baseline, for a fixed fidelity target across benchmarks.}
    \label{fig:combined_shots_vs_fidelity}
\end{figure*}

\subsection{Baseline and Evaluation Parameters}

The baseline is conventional VQA, where each task within an application is executed independently with an equal allocation of shots. Unless otherwise specified, we employ the SPSA optimizer~\cite{SPSA} with a mini-batch size ($N_{\text{evals-per-iter}}$) of 2.
For rigorous and consistent comparison, we set $N_{\text{per-eval}} = 4096 *$ (number of Pauli terms) for all VQA tasks in both the baseline and TreeVQA. Thus, each Pauli term is sampled 4096 times per evaluation, ensuring statistical accuracy in the measurement outcomes (see \autoref{sec:cost_exe_vqa}). As detailed in Section~\ref{sec:cost_exe_vqa}, this is a conservative estimate; for larger Hamiltonians, TreeVQA can yield even greater shot reductions. The total shot count is given by $N_{\text{overall}} = (\text{\# iterations}) \times 2 \times 4096 \times$ (number of Pauli terms). While $N_{\text{per-eval}}$ could be further reduced by grouping Pauli terms into Qubit-Wise Commuting (QWC) sets, this constant-factor improvement does not affect the relative shot reduction ratio, which is our primary metric.
Benchmark-specific shot budgets are chosen to ensure TreeVQA reaches convergence. Iteration counts typically range from 16,000 to 30,000, except for the simpler $H_2$ system, which converges within 300 iterations.

\subsection{Simulation Framework}
TreeVQA is evaluated using Qiskit's \texttt{AerSimulator}~\cite{qiskit2024}. For noiseless simulations, we utilize the \texttt{StatevectorSimulator} to obtain exact quantum state distributions. For realistic noise studies (see Section~\ref{subsec:noisy_execution}), we employ Qiskit's density matrix simulator with device-calibrated noise models, incorporating both discrete Pauli errors (from reset, measurement, and gate operations) and continuous decoherence processes (T1, T2). 
To enable scalable simulation of large systems, we leverage the PauliPropagation method~\cite{pauli_propagation}, which provides efficient, low-error simulation beyond the capabilities of conventional approaches, enabling studies of systems such as $\mathrm{C_2H_2}$ and 25-site Ising chains. For all noiseless experiments, we use the Hardware Efficient Ansatz~\cite{qiskit_efficientsu2} with two layers of circular entanglement; in noisy settings, we increase to five layers to amplify the effects of noise on fidelity.

\section{Evaluation}
\label{sec:evaluation}

\subsection{VQA Shot Reduction for a Fidelity Target}
\label{subsec:shot_reduction}

The goal of this evaluation is to estimate the VQE shot savings achieved by TreeVQA over the baseline, which executes each VQE task independently, for different fidelity thresholds. Results are shown in \autoref{fig:combined_shots_vs_fidelity}, comparing the total shots required by TreeVQA and the baseline across all tasks.

Across all six VQE benchmarks, TreeVQA consistently yields substantial shot savings. For instance, in the $HF$ molecule case, the baseline VQE attains about 98\% fidelity across tasks for $\sim 1.5 \times 10^{11}$ shots, whereas TreeVQA achieves the same fidelity with only $4 \times 10^9$ shots — a $34.7\times$ reduction.
As the target fidelity approaches $1.00$, TreeVQA's shot count rises because clustered tasks are eventually separated into distinct VQE instances to reach high-accuracy solutions, each requiring multiple iterations to converge. Importantly, the baseline approach would also consume these shots and would have already expended a substantial number to reach this fidelity, often making simulation impractical. Thus, TreeVQA remains more resource-efficient across the fidelity range.

Similar trends appear across other molecular and physical systems. Shot savings typically range from $30\times$ to $40\times$, though some cases differ. In the \(H_2\) UCCSD benchmark, the small problem size allows both methods to converge in fewer iterations, yielding smaller savings. In the XXZ model, slow convergence causes TreeVQA to split tasks aggressively, slightly reducing its relative advantage. These cases suggest that further refinements, such as more adaptive threshold tuning, could improve TreeVQA’s performance on more challenging optimization landscapes.

An interesting finding is that TreeVQA often surpasses the nominal shot-saving bound set by subproblem size. This arises from its effect on the SPSA learning rate,
$
\eta_k = \frac{a}{(A + k + 1)^\alpha},
$
where the superposition of Hamiltonians increases landscape curvature, driving larger $a$ and thus steeper gradients. The higher learning rates speed convergence and persist even when baselines are tuned, highlighting TreeVQA’s efficiency.

\begin{figure*}[htbp]
    \centering
    \includegraphics[width=\textwidth]{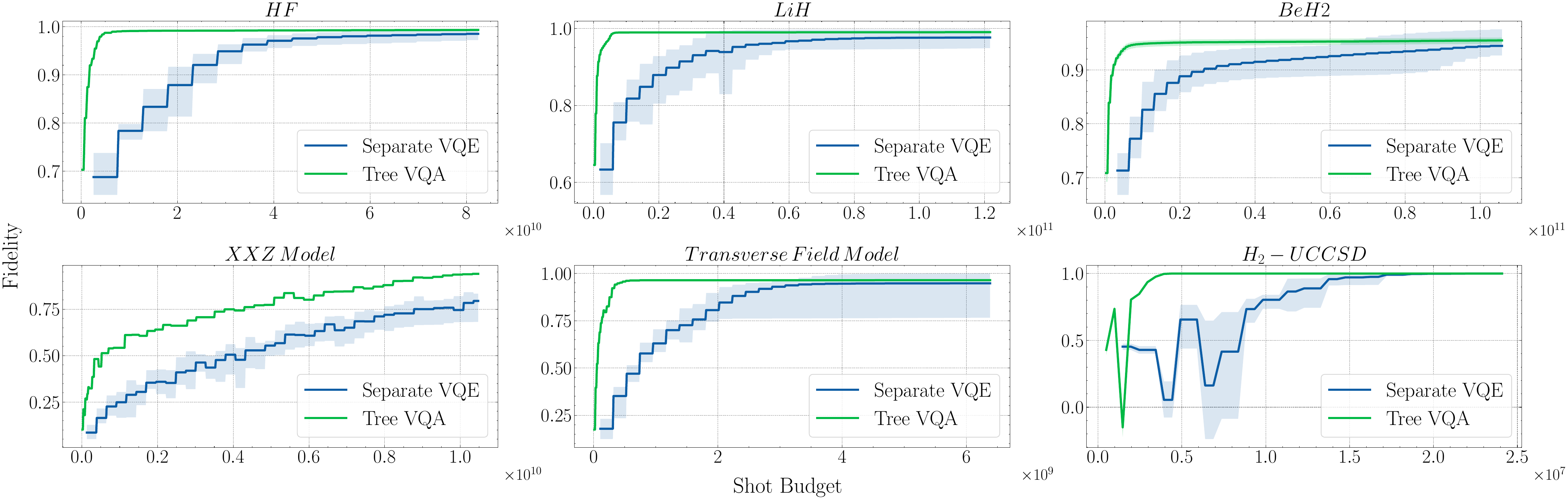}
    \caption{Fidelity gain achieved by TreeVQA compared to baseline execution for a fixed shot budget across all benchmarks}
    \label{fig:combined_fidelity_vs_shot_budget}
\end{figure*}

\begin{redtext}
\subsection{Fidelity Benefit for a Fixed Shot Budget}
\label{subsec:fidelity_benefit}
Another way to interpret TreeVQA’s benefit is through fidelity improvements under a fixed shot budget. As shown in ~\autoref{fig:combined_fidelity_vs_shot_budget}, TreeVQA consistently achieves higher fidelity compared to the baseline. 
Additionally, TreeVQA leads to lower fidelity variance in fidelity across different tasks (e.g., varying bond lengths or magnetic field strengths). This is because the tasks in TreeVQA are solved in a coordinated fashion, guided by shared information within clusters. In contrast, baseline VQEs, executed independently, often exhibit large disparities in fidelity under a uniform shot budget, since the optimal allocation for each task is unknown in advance.
\end{redtext}

\begin{figure}[t]
    \centering
    \includegraphics[width=\columnwidth,trim={0cm 0cm 0cm 1cm},clip]{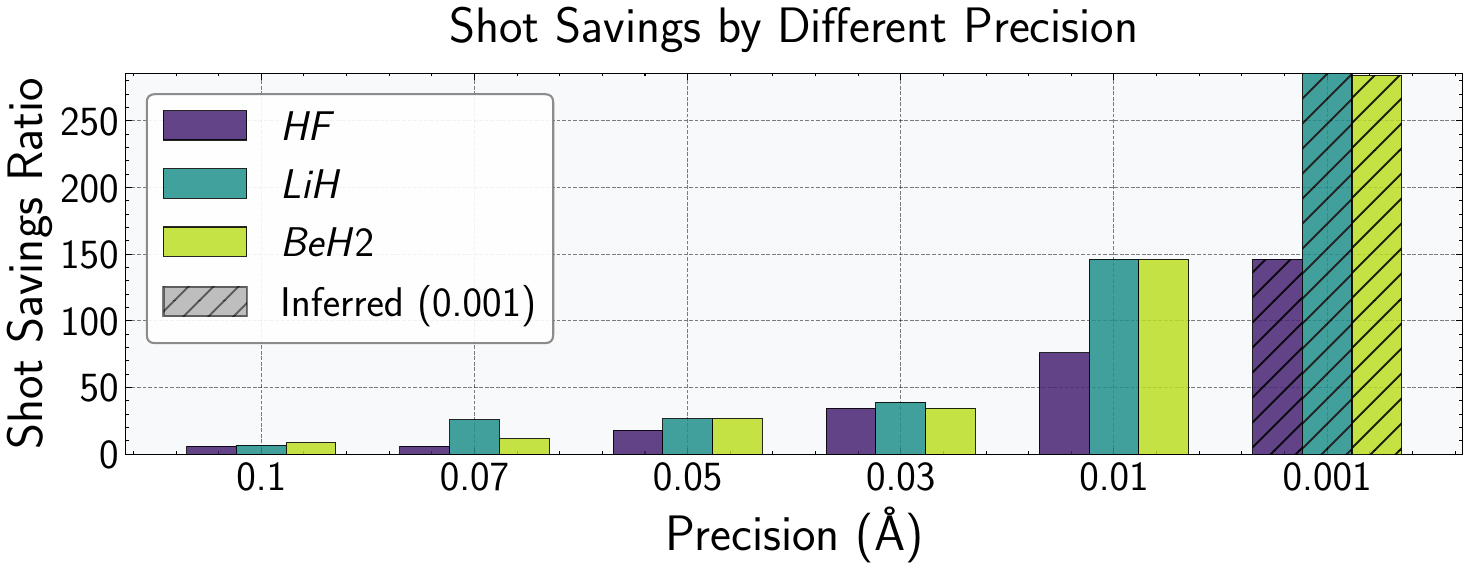}
    \caption{Shot savings at different precision levels}
    \label{fig:precision_shot_savings}
\end{figure}

\subsection{Benefits at Varying Task Precision}
\label{subsubsec:step_size}

Figure~\ref{fig:precision_shot_savings} shows that TreeVQA’s shot savings increase with higher precision requirements (smaller step sizes) over a fixed bond length range. As precision increases, the number of tasks grows (3, 5, 7, 10, and 30). Due to computational constraints, results for the finest step size ($0.001$) are extrapolated, indicated by shaded bars. Shot savings rise from $5$–$10\times$ at coarse precision to $80$–$100\times$ at finer precision, exceeding $250\times$ at the highest precision. This trend is intuitive: higher precision creates more subproblems with highly similar Hamiltonians, which can be effectively co-optimized within a single VQA cluster, maximizing quantum resource efficiency. Since practical applications often require high-precision energy estimates, TreeVQA is expected to provide substantial benefits in real-world scenarios.

\begin{figure}[t]
    \centering
    \includegraphics[width=\columnwidth]{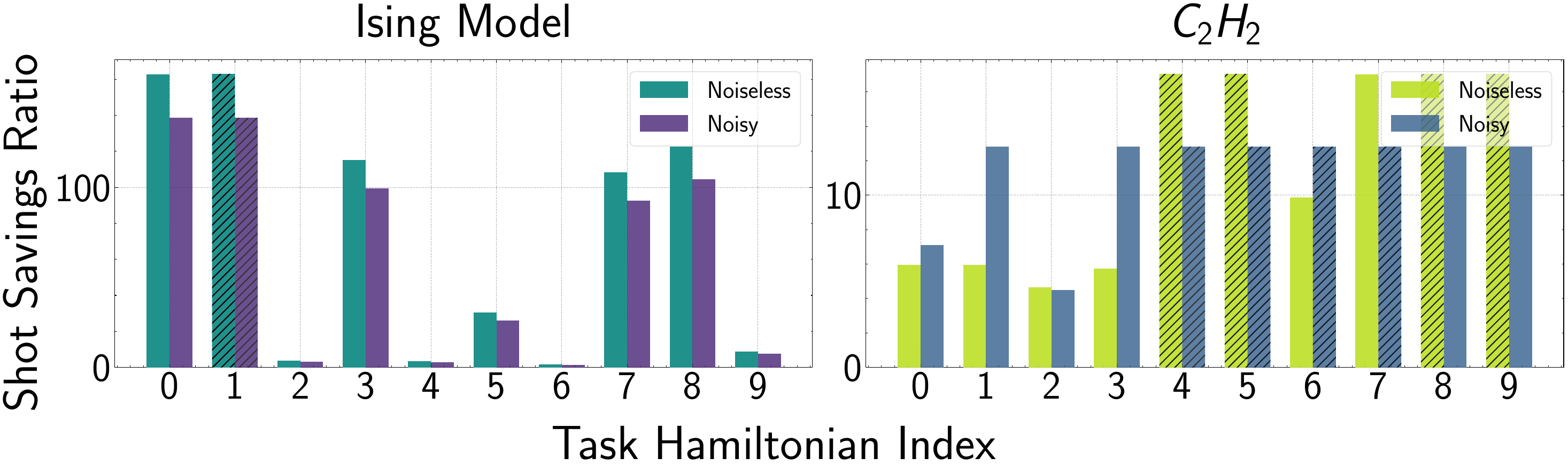}
    \caption{Shot savings on large-scale applications.}
    \label{fig:large_scale}
\end{figure}

\subsection{Estimating Benefits for Large-Scale Problems}
To evaluate TreeVQA on large-scale problems that may approach near-term quantum utility, we consider two benchmarks: $C_2H_2$ and a 25-site Ising model (each with ten tasks), using 28 and 50 qubits, respectively. Simulations are performed with the PauliPropagation method, truncating Pauli terms with frequencies above 8~\cite{pauli_propagation}. Results are shown in \autoref{fig:large_scale}. As classical solvers cannot provide exact ground states for these systems, we allocate a fixed number of iterations and measure the number of shots the baseline VQE requires to reach comparable energies. For some tasks, the baseline does not achieve TreeVQA’s energies within the given shots (indicated with hatch lines), suggesting that the actual shot savings will be even higher. Overall, TreeVQA demonstrates substantial shot reductions over the baseline across both benchmarks.
The lower benefit observed for $C_2H_2$ compared to the Ising model partly reflects the limited computational budget allocated to $C_2H_2$ (due to runtime constraints). In fact, the average energy gap between TreeVQA and the baseline is greater for $C_2H_2$ than for the Ising model. With additional budget, the shot savings for $C_2H_2$ would likely approach those observed for the Ising.
\begin{redtext}
Additionally, we evaluate TreeVQA under a noise model similar to that of~\cite{rudolph2025paulipropagation_error_mitigation_notebook}, in which a depolarizing noise layer (with error rates of $1\%$) is inserted after each circuit repetition. Under this noise model, the advantages of TreeVQA are marginally reduced relative to the noiseless case, yet it continues to deliver substantial shot savings compared to the baseline. Large-scale noise deforms the VQA optimization landscape and introduces additional local minima, adversely affecting both TreeVQA and the baseline and limiting their ability to converge to the ground state. This increased heterogeneity in the landscape slightly constrains TreeVQA’s ability to fully exploit task similarity. As quantum hardware advances in the EFT regime and noise levels continue to decrease, we expect TreeVQA’s performance gains to progressively approach those observed in noiseless simulations.

\end{redtext}

\begin{figure}[t]
    \centering
    \includegraphics[width=0.9\columnwidth,trim={0cm 0cm 0cm 0cm},clip]{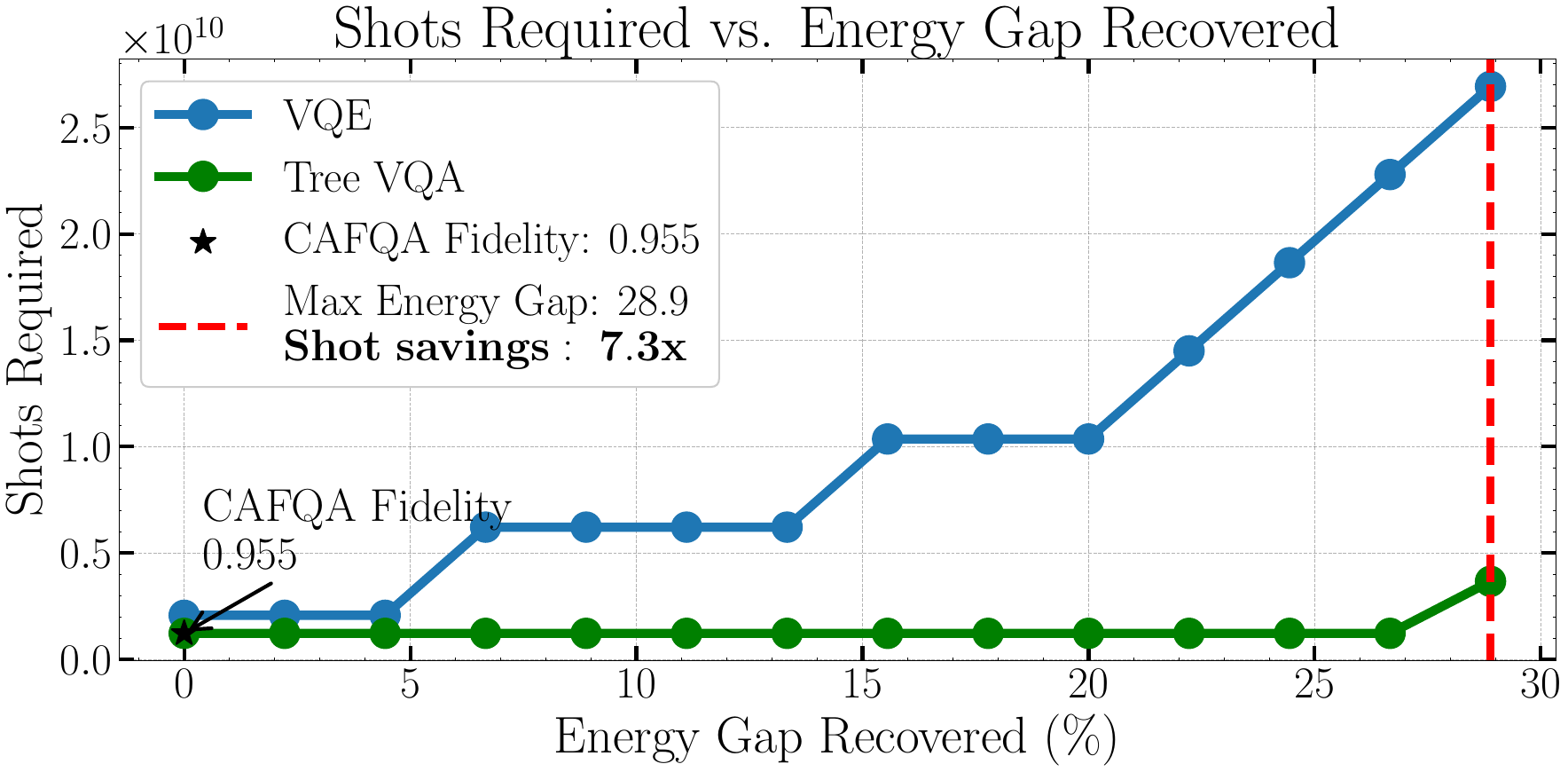}
    \caption{TreeVQA benefits for $LiH$ w/ CAFQA.}
    \label{fig:cafqa_LiH}
\end{figure}

\subsection{TreeVQA integrated with good VQE Initialization}
\label{subsec:cafqa}
Although TreeVQA significantly reduces quantum hardware execution (shot) costs for VQAs, it is not the only approach to do so. Shot costs can also be reduced via classical initialization: if good VQA parameters are identified classically, quantum execution can start from these parameters, accelerating convergence. For VQE, the state-of-the-art classical initialization method is CAFQA~\cite{CAFQA_Ravi2022} and its related work~\cite{seifert2024clapton,bhattacharyya2025excitedcafqaclassicalsimulationbootstrap,bhattacharyya2023optimal}, which find initial parameters using a Clifford-only search strategy. While CAFQA achieves high initialization accuracy, its benefits (and those of any classical initialization) diminish as the inherent \emph{quantum-ness} of the problem increases, since classical methods can only resolve the ``easy'' classical component, leaving the harder quantum portion to iterative quantum execution.

Even with accurate classical initialization, TreeVQA provides substantial additional benefits. In Fig.~\ref{fig:cafqa_LiH}, we evaluate the LiH molecule at a bond length of $3.15-3.24$ at a precision of 0.01 $\si{\angstrom}$, where CAFQA achieves 95.5\% initialization accuracy. Using CAFQA parameters to initialize both baseline VQE and TreeVQA, we find that TreeVQA recovers the residual energy gap between the CAFQA initialization and the true ground state with far fewer shots. Specifically, TreeVQA recovers 30\% of this gap using $7.3\times$ (i.e. $1.8\times 10^{10}$) fewer shots. These benefits are expected to increase for more challenging problems where classical initialization is far less effective.

\begin{figure}[htbp]
    \centering
    \includegraphics[width=\columnwidth]{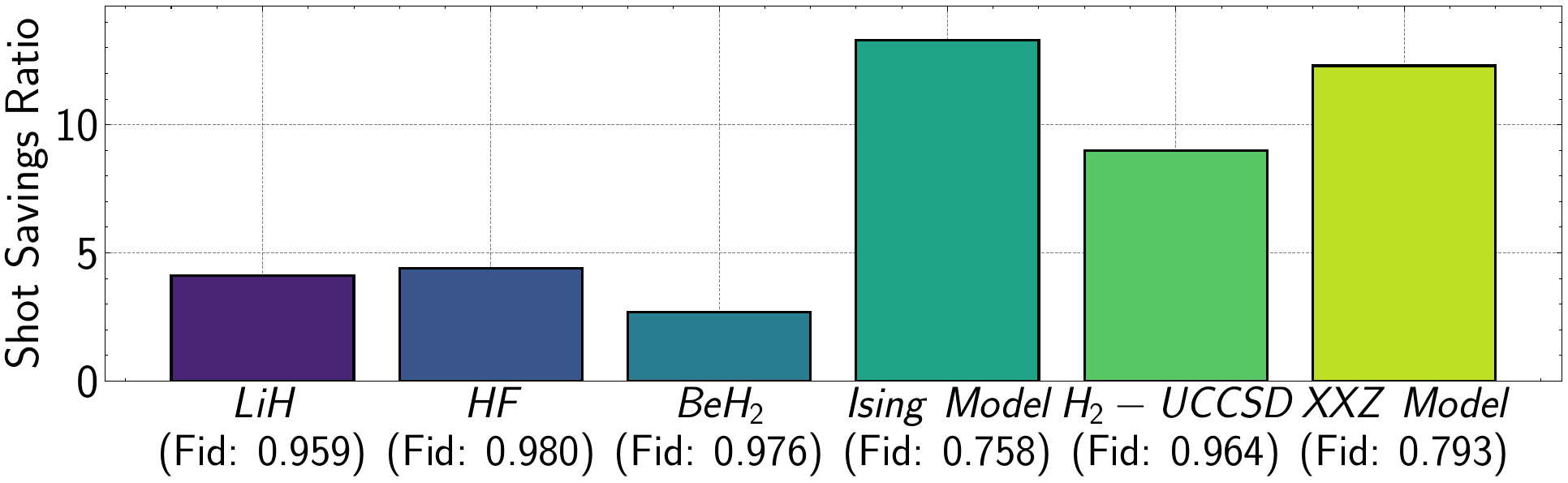}
    \caption{TreeVQA benefits for VQA w/ COBYLA optimizer.}
    \label{fig:cobyla_shots}
\end{figure}

\subsection{Untuned TreeVQA with Alternate Optimizer}

In previous sections, we demonstrated the benefits of TreeVQA with the SPSA optimizer, one of the most widely used optimizers for VQAs. To maximize performance with SPSA, we carefully tuned TreeVQA’s strategies and hyperparameters to align with SPSA’s functionality (see Section~\ref{subsubsec:window_size} and Section~\ref{subsubsec:threshold}). However, TreeVQA is not tied to any single optimizer. In this section, we highlight its effectiveness with an alternate optimizer, COBYLA (Constrained Optimization BY Linear Approximations), when running VQE across diverse applications. COBYLA can be advantageous in noisy, non-smooth optimization landscapes since it does not rely on gradients, but it scales poorly with large parameter counts and can become trapped in local minima. Unlike SPSA’s stochastic gradient-based updates, COBYLA relies on local linear approximations, making its optimization dynamics fundamentally different, potentially posing a challenge for TreeVQA. Thus, demonstrating TreeVQA’s benefits in this setting, without any additional fine-tuning, underscores its broad applicability. Remarkably, TreeVQA still achieves $2.5\times$--$13\times$ shot savings over traditional VQE with COBYLA, clearly showcasing its plug-and-play utility across optimizers.

\begin{redtext}
    
\begin{table}[hbp]
    \centering
    \caption{LiH TreeVQA Noisy Simulation Results}
    \small
    \begin{tabular}{|c|c|c|}
        \hline
        Backend & Max Avg & Shots Saving \\
        Name & Fidelity & Ratio \\
        \hline
        $Hanoi$ & $0.951$ & $12.0\times$ \\
        \hline
        $Cairo$ & $0.958$ & $17.0\times$ \\
        \hline
        $Mumbai$ & $0.910$ & $12.1\times$ \\
        \hline
        $Kolkata$ & $0.878$ & $24.8\times$ \\
        \hline
        $Auckland$ & $0.888$ & $14.7\times$ \\
        \hline
    \end{tabular}
    \label{tab:lih_tree_vqa_noise}
\end{table}
\end{redtext}
\subsection{Benefits in Noisy Execution Settings}
\label{subsec:noisy_execution}
All prior evaluations in this work are based on noiseless simulations. Since TreeVQA primarily modifies the VQA execution strategy rather than the hardware itself, benefits observed in noiseless settings are expected to largely carry over to real quantum devices. To validate this, we evaluate TreeVQA using noisy device simulation with $LiH$ as a benchmark. To accentuate noise effects, we increase the ansatz entanglement layers from 2 to 5, as deeper circuits are both more sensitive to noise and more representative of problem-tailored ansatz, which typically exceed the depth of the Hardware Efficient Ansatz~\cite{LUCJ}. 

For noisy simulations, we employ Qiskit's highest optimization level with error mitigation enabled~\cite{li2019sabre,murali2019noise}. The COBYLA optimizer is used instead of SPSA, as SPSA converges too slowly under noise. As shown in \autoref{tab:lih_tree_vqa_noise}, TreeVQA achieves substantial shot savings even in noisy conditions.

\subsection{TreeVQA Benefits for QAOA}
\label{subsec:qaoa}

\begin{figure}[t]
    \centering
    \includegraphics[width=\columnwidth]{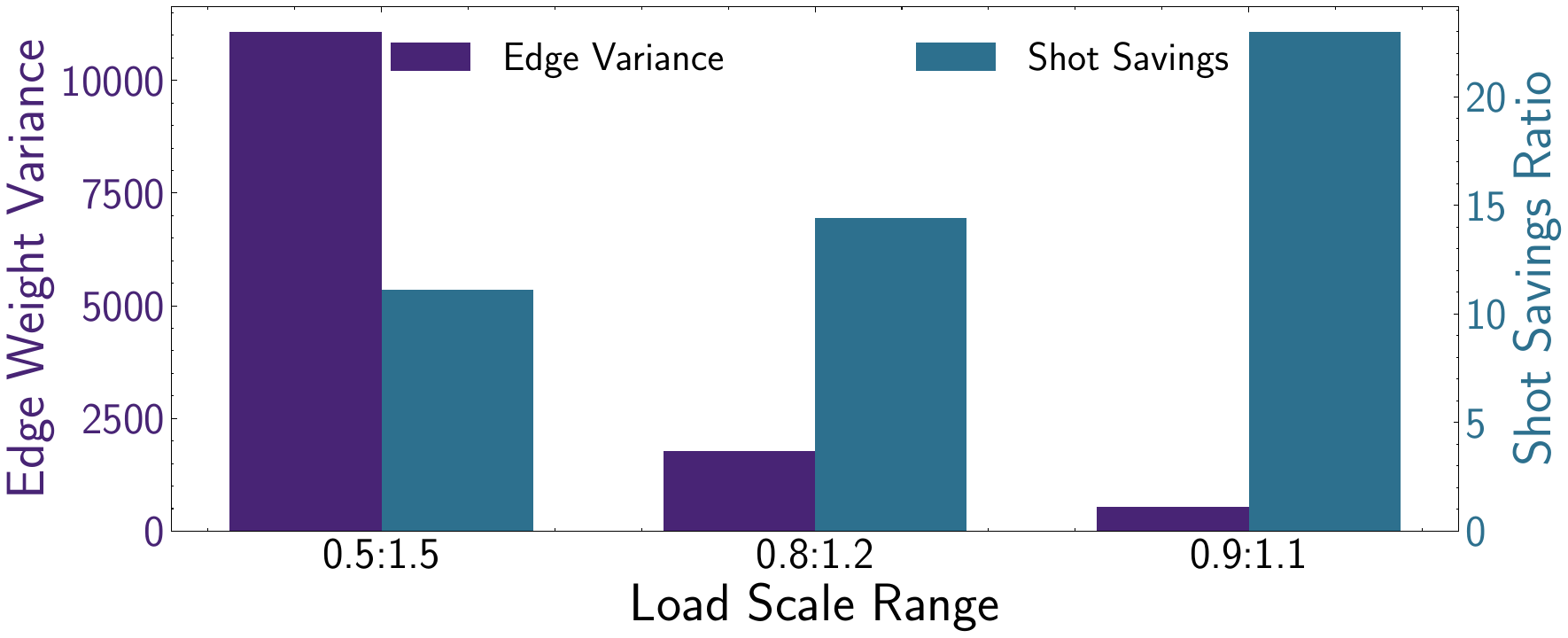}
    \vspace{-0.35cm}
    \caption{TreeVQA shot savings for QAOA.}
    \label{fig:qaoa_variance}
\end{figure}

To demonstrate TreeVQA's applicability beyond VQE, we evaluate its performance on QAOA for MaxCut problems using the IEEE-14 bus system (see \autoref{subsubsec:qaoa_benchmark}). This system is modeled as a 14-node undirected graph. We consider three load scale ranges: 0.5 to 1.5, 0.8 to 1.2, and 0.9 to 1.1, representing increasing levels of similarity among problem instances, from extreme planning scenarios to typical operational variations and small forecasting errors. For each range, we generate 10 graphs with edge weights determined by equally spaced load scales, and solve all 10 MaxCut instances jointly with a single TreeVQA run.
For all graphs in our experiments, we use the state-of-the-art classical QAOA initialization method Red-QAOA~\cite{red-qaoa}. Red-QAOA applies a graph-pooling technique to generate the initial state for unweighted MaxCut problems. In our case, since the graphs are isomorphic and differ only in edge weights, the initial state remains the same for all.

To quantify graph similarity within each range, we compute the edge weight variance (purple bars in \autoref{fig:qaoa_variance}). All graphs are aligned to a common node ordering, and the variance is calculated as the average squared deviation of each graph’s edge weights from the mean graph.
As shown in \autoref{fig:qaoa_variance}, TreeVQA achieves over $20\times$ shot savings (blue bars) when problem instances are more similar (lower variance), highlighting its ability to exploit shared structure across related tasks. Even as the variance increases, benefits remain significant (greater than $10\times$). These results confirm that TreeVQA’s benefits extend to combinatorial optimization and that leveraging task similarity can significantly reduce quantum resource requirements.

\section{Discussion}\label{sec:discussion}

\begin{redtext}
\subsection{Hyperparameter Analysis}
We examine two key TreeVQA hyperparameters—sliding window size and threshold—both of which will eventually affect the splitting timing. Thus we will first analyze the splitting timing effect. To further help the analysis, we introduce the \emph{Tree Critical Depth}, defined as the longest path from the root to a leaf in the TreeVQA tree (see \autoref{fig:treevqa_overview}). This metric serves as a proxy for splitting time, with more effective splitting typically resulting in shallower trees.

\textbf{Analysis of Cluster Splitting Timing:}
\label{subsec:split_analysis}
Cluster splitting timing is crucial to TreeVQA's performance, so we investigate how it impacts VQE accuracy and efficiency. We conduct experiments on three chemistry benchmarks, manually enforcing a single split during optimization; each point in \autoref{fig:combined_splitting} reports the final error when this split is scheduled at different stages of optimization (expressed as a percentage of total iterations). We found that the optimal timing is application-dependent (motivating an intelligent controller) but typically occurs mid-optimization (the $H_2$ benchmark performs best with later splitting due to its smaller problem size, which reduces the risk of overfitting even in later stages). Splitting too early prevents subproblems from exploiting shared parameter regions, wasting resources, while splitting too late causes overfitting to the mixed Hamiltonian, requiring extra shots to escape local minima. Both cases yield fidelity losses of several percentage points under a fixed shot budget, large enough to compromise chemical accuracy.

\begin{figure}[t]
    \centering
    \includegraphics[width=\linewidth]{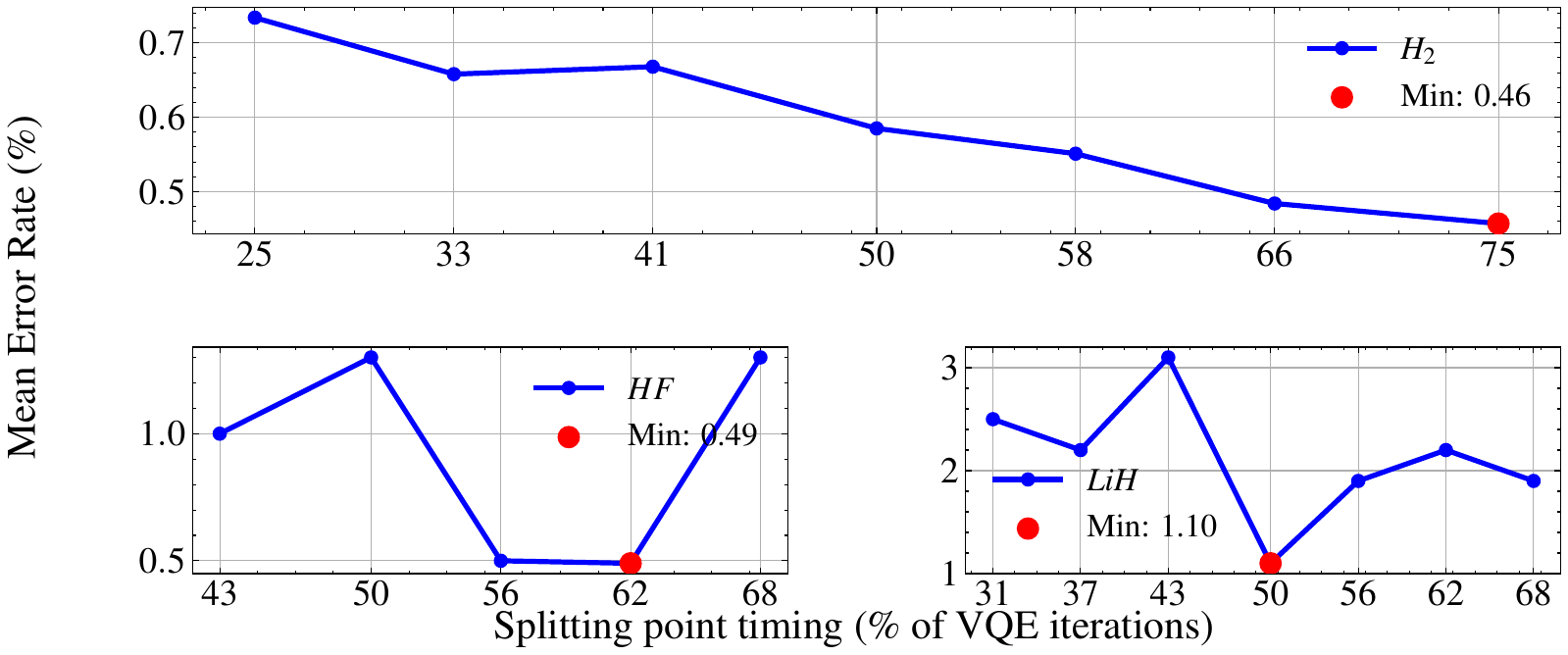}
        \vspace{-0.5cm}
    \caption{Analysis of splitting points and their impact on TreeVQA performance}
    \label{fig:combined_splitting}
\end{figure}

\textbf{Window Size Analysis:}
\label{subsubsec:window_size}
The window size determines how many iterations are averaged to estimate the gradient slope for convergence detection (Sec.~\ref{sec:optimization}). \autoref{fig:window_plot} shows how this hyperparameter affects TreeVQA's performance. Small windows make the algorithm noise-sensitive, causing premature splitting, while overly large windows delay needed splits and risk overfitting. Sweep experiments show that a moderate window of $0.01\%$–$0.02\%$ of total iterations strikes the best balance, capturing convergence patterns while remaining responsive. This empirically yields up to a $5\times$ reduction in VQA error compared to other choices. Adaptive windowing may offer further gains.

\textbf{Splitting Threshold Analysis:}
\label{subsubsec:threshold}
The threshold parameter controls the aggressiveness of splitting, with larger values making splits less frequent. Sweeping a logarithmic range of thresholds reveals an optimal middle ground, achieving up to a $5\times$ reduction in overall VQA error under a fixed shot budget. Excessively high thresholds delay splitting, causing overfitting to the mixed Hamiltonian, while low thresholds trigger premature splits and waste shots. In practice, suitable thresholds can be selected by profiling gradient slopes, as optimal values depend on Hamiltonian structure and optimizer dynamics. Overall, even with moderate hyperparameter choices, TreeVQA consistently provides significant shot savings and high accuracy.
%, demonstrating robustness compared to baseline VQE.

\begin{figure}[t]
  \centering
  \includegraphics[width=\columnwidth]{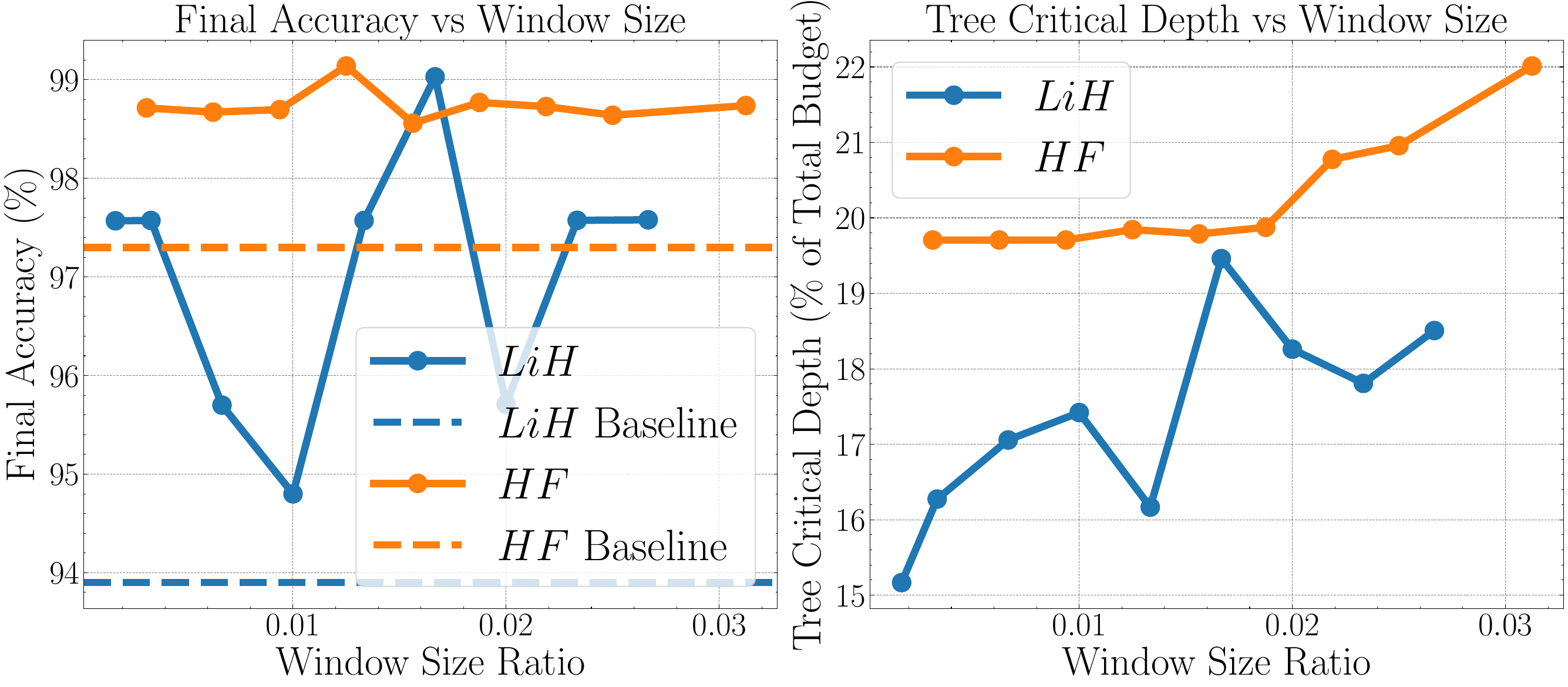}
  %\vspace{-0.2cm}
  \caption{Window size analysis}
  \label{fig:window_plot}
\end{figure}

\end{redtext}

\subsection{Expanding TreeVQA Scope}
\textbf{Beyond the Target Applications:}
TreeVQA’s hierarchical clustering and adaptive shot allocation naturally extend to Quantum Machine Learning (QML). Penalty terms in QML loss functions can play roles analogous to bond lengths or magnetic fields, enabling efficient training of multiple models on the same dataset with reduced quantum costs.

\textbf{Beyond SPSA and COBYLA:}
TreeVQA is compatible with any optimizer, requiring only cost function evaluations. It can also leverage optimizer-specific behavior, for instance, using Implicit Filtering’s adaptive step sizes to adjust cluster granularity dynamically: coarser grouping during broad exploration and finer clustering when high precision is needed.

\begin{redtext}

\textbf{Beyond the NISQ Era:}
Although VQAs are often associated with NISQ due to their noise resilience, they remain compelling because of their practical applications and strong theoretical support~\cite{zoltan2025myths}. VQAs are also relevant in the Early Fault-Tolerant (EFT) regime, where limited quantum error correction still yields logical error rates too high for algorithms like QPE or Shor's algorithm. Consequently, recent efforts have extended NISQ-like methods into the EFT regime, including work on uncorrected RZ gates with error-corrected Cliffords~\cite{ismail2025transversalstararchitecturemegaquopscale} and VQAs on partially error-corrected devices~\cite{dangwal2025variationalquantumalgorithmsera,eft}. Since TreeVQA operates at a higher level of abstraction, it can be applied to EFT systems regardless of the underlying gate set (e.g., Clifford+T or Clifford+RZ), enabling significant resource savings in this emerging regime. 

\section{Related Work}
\label{sec:related_work}
Measurement reduction techniques often group Pauli strings that are either mutually qubit-wise commuting~\cite{gokhale2020on3} or transformed into compatible unitary sub-Hamiltonians~\cite{yen2020measuring}. Others, like ADAPT-VQE~\cite{adaptvqe}, truncate terms based on energy contributions. While both orthogonal and compatible with TreeVQA, these approaches introduce trade-offs, such as complex grouping heuristics or increased circuit depth from additional unitaries. Similarly, iteration reduction methods—using classical surrogates for warm-starts~\cite{CAFQA_Ravi2022} or problem simplification~\cite{red-qaoa}—seamlessly integrate with our work (Sec.~\ref{subsec:cafqa},~\ref{subsec:qaoa}) but fail to address inherent multi-task structures. To our knowledge, Meta-VQE~\cite{Cervera2021} remains the only prior work explicitly targeting multi-task VQE by embedding Hamiltonians parameters directly into a specialized ansatz, it lacks the dynamic monitoring and adaptive splitting mechanisms required to achieve the substantial shot savings offered by TreeVQA.
\end{redtext}

\section{Conclusion}\label{sec:conclusion}

We introduced TreeVQA, a hierarchical framework that significantly reduces quantum resource consumption for variational quantum algorithms (VQAs). By dynamically clustering similar Hamiltonians and adaptively allocating shots, TreeVQA achieves orders-of-magnitude measurement savings without compromising accuracy.
This approach addresses the core challenge of costly quantum resources, bringing VQAs closer to practical applications. As quantum hardware advances, TreeVQA’s scalable design remains relevant, supporting efficient execution in future HPC+Quantum systems and fault-tolerant architectures.

\section{Acknowledgment}\label{sec:acknowledgment}

This material is based upon work supported by the U.S. Department of Energy, Office of Science, Office of Advanced Scientific Computing Research, Accelerated Research in Quantum Computing under Award Number DE-SC0025633. This research used resources of the National Energy Research Scientific Computing Center, a DOE Office of Science User Facility supported by the Office of Science of the U.S. Department of Energy under Contract No. DE-AC02-05CH11231 using NERSC award ASCR-ERCAP0033197.

%\input{./sections/10_Appendix}
%%%%%%%%%%%%%%%%%%%%%%%%%%%%

% use the ACM bibliography style
\bibliographystyle{ACM-Reference-Format}
\bibliography{references,ref1}

\end{document}